\documentclass[%
reprint,
 amsmath,amssymb,
 aip,
]{revtex4-2}

\usepackage{graphicx}
\usepackage{dcolumn}
\usepackage{bm}
\usepackage{afterpage}
\usepackage{amssymb}
\usepackage{graphicx}
\usepackage[version=4]{mhchem}
\usepackage[export]{adjustbox}
\usepackage{xcolor}
\usepackage{tabularx}
\usepackage{textgreek}
\usepackage{float}
\usepackage[section]{placeins}
\usepackage{epstopdf}
\usepackage{textcase}

\begin{document}


\title{Jahn-Teller Distortion and Dissociation of \ce{CCl4+} by Transient X-ray Spectroscopy Simultaneously at the Carbon K- and Chlorine L-Edge}
\author{Andrew D. Ross}
\altaffiliation{These authors contributed equally to this work.}
\affiliation{%
 Department of Chemistry, University of California, Berkeley, CA, 94720, USA
}
\affiliation{%
 Chemical Sciences Division, Lawrence Berkeley National Laboratory, Berkeley, CA, 94720, USA
}

\author{Diptarka Hait}%
\altaffiliation{These authors contributed equally to this work.}
\affiliation{%
 Department of Chemistry, University of California, Berkeley, CA, 94720, USA
}
\affiliation{%
 Chemical Sciences Division, Lawrence Berkeley National Laboratory, Berkeley, CA, 94720, USA
}
\author{Valeriu Scutelnic}%
\affiliation{%
 Department of Chemistry, University of California, Berkeley, CA, 94720, USA
}
\affiliation{%
 Chemical Sciences Division, Lawrence Berkeley National Laboratory, Berkeley, CA, 94720, USA
}
\author{Eric A. Haugen}%
\affiliation{%
 Department of Chemistry, University of California, Berkeley, CA, 94720, USA
}
\affiliation{%
 Chemical Sciences Division, Lawrence Berkeley National Laboratory, Berkeley, CA, 94720, USA
}
\author{Enrico Ridente}%
\affiliation{%
 Department of Chemistry, University of California, Berkeley, CA, 94720, USA
}
\author{Mikias B. Balkew}%
\affiliation{%
 School of Physics, Georgia Institute of Technology, Atlanta, GA, 30332, USA
}
\affiliation{%
 Department of Chemistry, University of California, Berkeley, CA, 94720, USA
}

\author{Daniel M. Neumark}%
\affiliation{%
 Department of Chemistry, University of California, Berkeley, CA, 94720, USA
}
\affiliation{%
 Chemical Sciences Division, Lawrence Berkeley National Laboratory, Berkeley, CA, 94720, USA
}
\author{Martin Head-Gordon}%
\affiliation{%
 Department of Chemistry, University of California, Berkeley, CA, 94720, USA
}
\affiliation{%
 Chemical Sciences Division, Lawrence Berkeley National Laboratory, Berkeley, CA, 94720, USA
}

\author{Stephen R. Leone}%
\affiliation{%
 Department of Chemistry, 
 University of California, Berkeley, CA, 94720, USA
}
\affiliation{%
 Chemical Sciences Division, 
 Lawrence Berkeley National Laboratory, Berkeley, CA, 94720, USA
}
\affiliation{%
 Department of Physics, 
 University of California, Berkeley, CA, 94720, USA
}%


\date{\today}

\begin{abstract}

X-ray Transient Absorption Spectroscopy (XTAS) and theoretical calculations are used to study \ce{CCl4+} prepared by 800 nm strong-field ionization. XTAS simultaneously probes atoms at the carbon K-edge (280-300 eV) and chlorine L-edge (195-220 eV). Comparison of experiment to X-ray spectra computed by orbital-optimized density functional theory (OO-DFT) indicates 
that after ionization, \ce{CCl4+} undergoes symmetry breaking driven by Jahn-Teller distortion away from the initial tetrahedral structure (T$_d$) in 6$\pm$2 fs. The resultant symmetry-broken covalently bonded form subsequently separates to a noncovalently bound complex between \ce{CCl3+} and \ce{Cl} over 90$\pm$10 fs, which is again predicted by theory. Finally, after more than 800 fs, L-edge signals for atomic Cl are observed, indicating dissociation to free \ce{CCl3+} and \ce{Cl}. The results for Jahn-Teller distortion to the symmetry-broken form of \ce{CCl4+} and formation of the \ce{Cl---CCl3+} complex characterize previously unobserved new species along the route to dissociation. 

\end{abstract}


\maketitle

\section{\label{sec:level1a}Introduction}
The Jahn-Teller (JT) theorem\cite{jahn1937stability} states that degenerate electronic states in non-linear molecules cannot be minima of energy vs nuclear positions, and will therefore undergo nuclear displacements that break the degeneracy. This is often observed in highly symmetric molecules 
with partial occupation of symmetry equivalent orbitals. 
A prototypical example of JT distortion is \ce{CH4^+}, which arises from ionization from the triply degenerate highest occupied molecular orbital (HOMO) of \ce{CH4}. The resulting cation rapidly distorts away from the tetrahedral (T$_d$)  geometry of the neutral to a C$_{2v}$ symmetry structure\cite{Knight1984MethaneJTObserve,Boyd1991MethaneJTEnergy} with two long and two short \ce{CH} bonds.
Baker et al.\cite{baker2006probing} have attempted to measure the time required for the JT process via comparison of high harmonic generation in \ce{CH4+} to \ce{CD4+}, where they report a difference between the geometry relaxation times of the two molecules up to 1.6 fs \textcolor{black}{with sub-cycle resolution. However, information about longer times was limited by the wavelength of the laser}. Recently, Gon\c{c}alves et al.\cite{Remacle2021CH4JT} calculated the JT timescale in \ce{CH4+} and found that distortion to C$_{2v}$ occurs as early as 7 fs, while the heavier \ce{CD4+} takes longer to relax to the same structure.

Ionization of halogenated methanes is expected to lead to slower (and thus easier to measure) JT distortion, on account of the larger substituent masses. 
However, JT distortion has not been observed for such systems, as the resulting cations are quite short-lived and therefore difficult to study\cite{Benyon1960MassSpecCCl4}. 
Indeed, it has been assumed that dissociation of the T$_d$ conformation of \ce{CCl4^+} to \ce{CCl3^+} and \ce{Cl} occurs without any intermediates\cite{Tachikawa1997CCl4Dissociation, harvey2014photoionization}, as no stable ion is produced by direct photoionization\cite{Kinugawa2002photoionization}. Theoretical work\cite{GarciaDeLaVegaDFT1997,harvey2014photoionization} has suggested a stable structure of the form \ce{[Cl3C---Cl]^+}, with one Cl 3.5 {\AA} away from the C and weakly coordinating to it. However, there are no experiments that validate the existence of these possible \ce{CCl4+} complexes.

\begin{figure*}[!t]
\includegraphics[max size={\textwidth}{\textheight/2}]{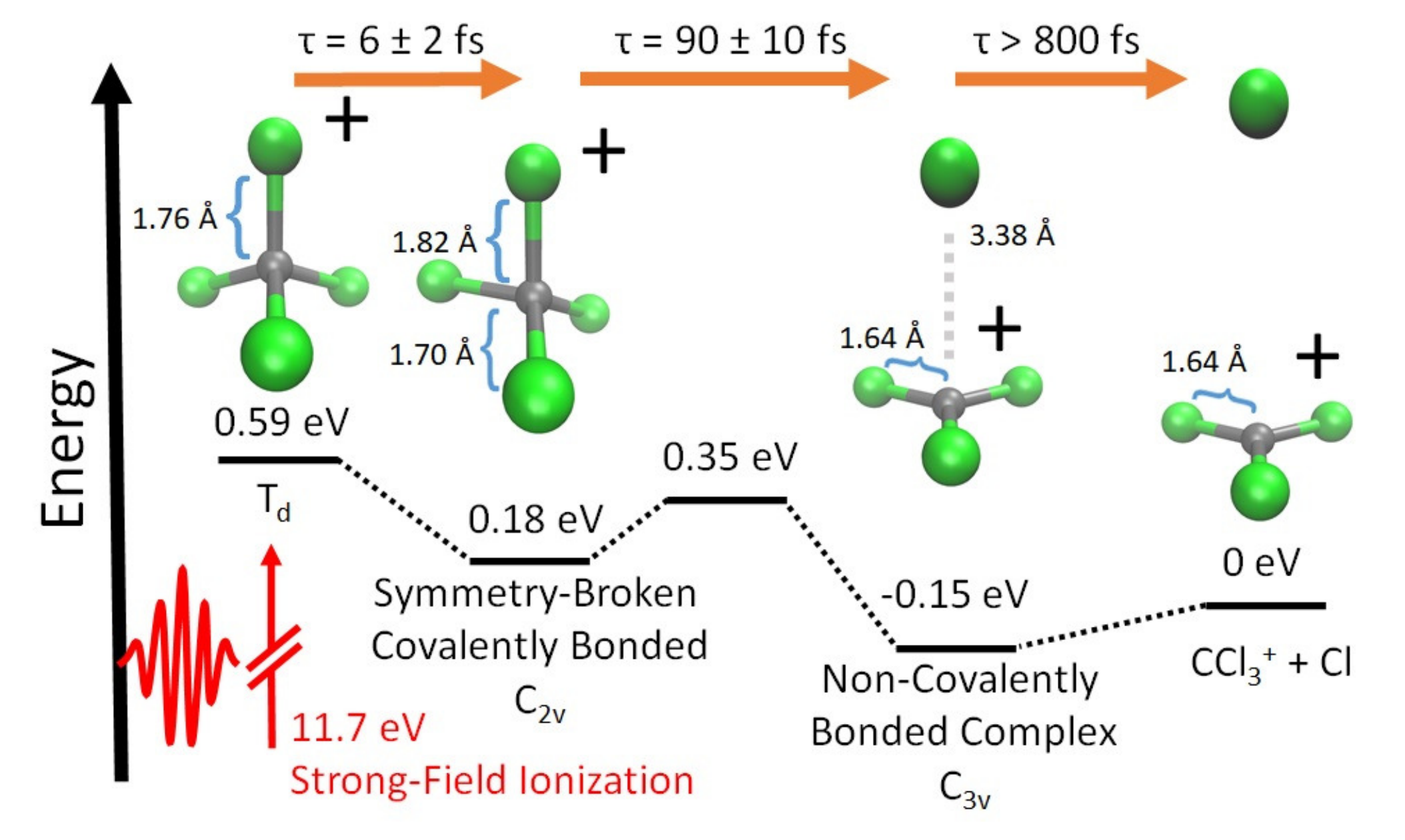}
\caption{\label{fig:Fig. ExperimentExplanation} 
The sequence of computationally characterized intermediates for the dissociation of \ce{CCl4+} identified in this experiment. Geometries were optimized with $\omega$B97M-V\cite{wB97MV}/aug-pcseg-3\cite{jensen2014unifying}. Zero-point energies were computed at the same level of theory. Electronic energies at these geometries were subsequently found using CCSD(T)\cite{raghavachari1989fifth} extrapolated to the complete basis set (CBS) limit. The structures of the closest local minima are used for the symmetry-broken covalently bonded form and non-covalently bonded complex; although, the experiment samples a large range of nuclear configurations, due to vibrational energy. The energy is set to 0 for the dissociated ion. Time constants shown are gathered from the $\Delta$OD data in Fig. \ref{fig:deltaODFig}b and c and their lineouts \textcolor{black}{(as described in the electronic supporting information [ESI])}. \textcolor{black}{We refer to the symmetry-broken covalently bonded form as SBCB and non-covalently bonded complex as NBC, in subsequent figures and tables.}
}

\end{figure*}

The instability of \ce{CCl4+} necessitates the use of a technique that is sensitive to changes in nuclear and electronic structure occurring within the first few femtoseconds after formation. X-ray Transient Absorption Spectroscopy (XTAS) is such a method, as its time resolution is only limited by the duration of the pump and probe laser pulses and femtosecond resolution is readily achievable\cite{Barreau2020,chang2010attosecond,li53asXRay}. Following strong-field ionization by the pump, the X-ray pulse induces transitions from core levels to unoccupied orbitals, which ensures that XTAS is sensitive to the local environment of individual atoms in the probed molecular system\cite{chang2010attosecond,Geneaux2019ATASReview}. It is therefore well suited for probing possible JT distorted transient intermediates that may arise during a dissociation. Indeed, XTAS has previously been used to infer JT distortion in the benzene cation\cite{Ephshtein2020BenzenCation} and ring opening in cyclohexadiene\cite{Attar2017RingOpening}. Pertot et al.\cite{Worner2017CF4} also employed XTAS at the C K-edge to study the dissociation of \ce{CF4+}, reporting rapid dissociation to \ce{CF3+} and \ce{F} within $\sim$40 fs. 

While several aforementioned studies focused on the C K-edge, HHG X-ray pulses span a relatively broad energy range\cite{Barreau2020,teichmann20160,johnson2018high} and can be used to study multiple atomic edges simultaneously. In particular, the Cl L$_{2,3}$-edge (2\textit{p} levels, $\sim$195-220 eV) is a natural complement to the C K-edge (1\textit{s} level, $\sim$280-300 eV) for studying chlorinated methanes.  Signal from two elements potentially allows for the identification of species whose spectra might be unresolvable or involve forbidden transitions at any one particular edge. It also permits observation of any Cl dissociation both from the perspective of the departing atomic Cl species and the remaining C-containing molecular fragment. 


\begin{figure*}[!t]
\includegraphics[max size={\textwidth}{0.6\textheight}]{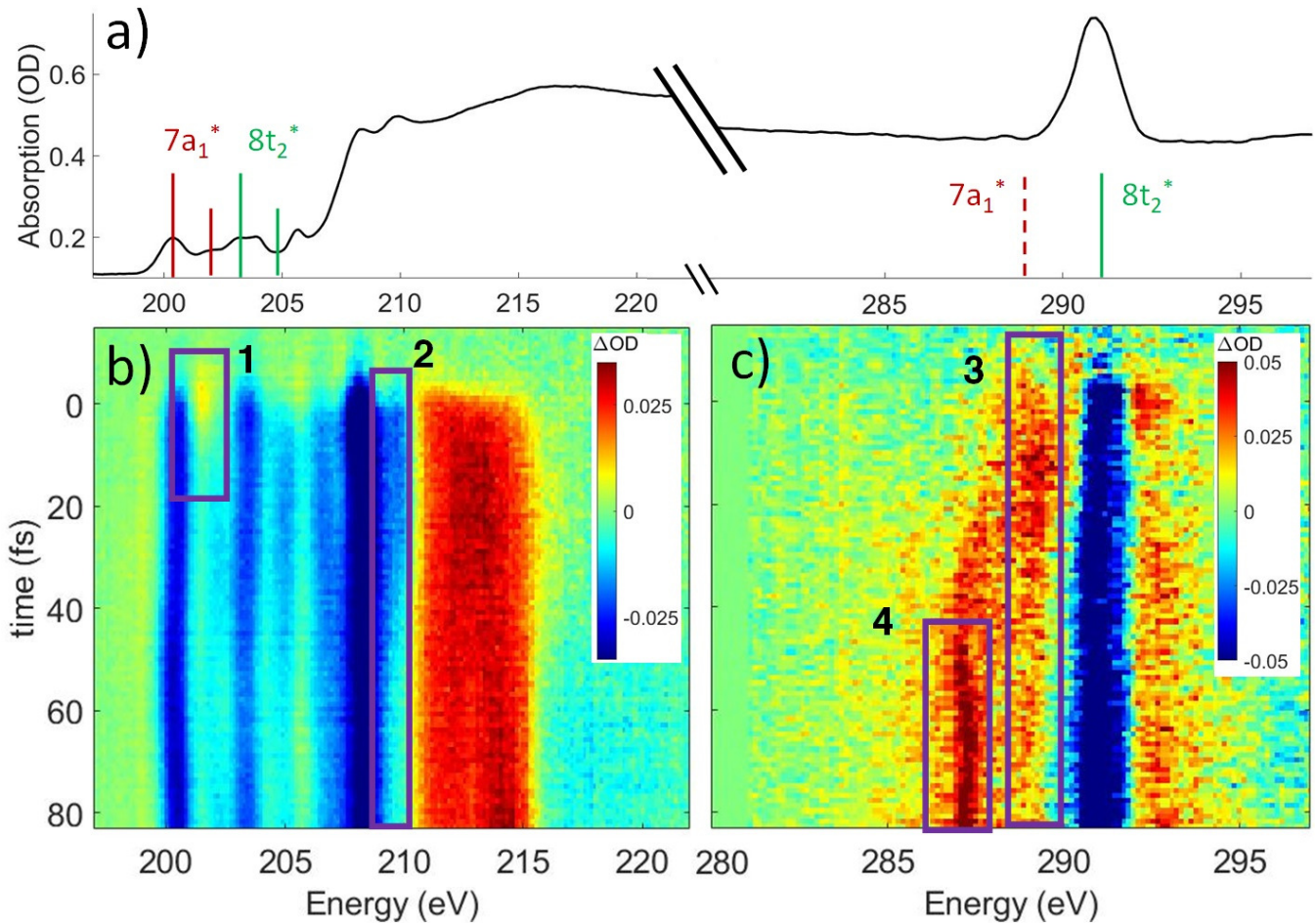}
\caption{\label{fig:deltaODFig} a: Static absorption spectrum of neutral \ce{CCl4} at the Cl L$_{2,3}$- and C K-edges, dominated mostly by transitions to the singly degenerate $7a_1^*$ and triply degenerate $8t_2^*$ levels. The C $1s\to 7a_1^*$ transition is dipole forbidden, and  absent in the static spectrum. b \& c: $\Delta$OD data from the highest pump power ($\sim$3$\times$10$^{14}$ W/cm$^2$) for the Cl L$_{2,3}$-edge (b) and C K-edge (c). 
Positive time corresponds to 800 nm pump first, while negative time is X-ray probe first. Negative $\Delta$OD represents depletion of neutral \ce{CCl4}, while positive $\Delta$OD features indicate presence of new species that absorb stronger than the parent in that region. Prominent transient features are labeled 1 (201.7 eV), 2 (209.5 eV), 3 (289.2 eV) and 4 (287.1 eV). 
A discussion of the features and their assignments is presented in the text. Additional data sets are shown in the SI.
}
\end{figure*}

In this paper, we present the results of a time-resolved experimental and theoretical investigation of strong-field ionized \ce{CCl4}, by tabletop XTAS simultaneously at both the C K-edge and the Cl L$_{2,3}$-edge. We report evidence for a dissociation pathway of \ce{CCl4+}, illustrated in Fig. \ref{fig:Fig. ExperimentExplanation}, that is initiated by an ultrafast ($\sim$6 fs) JT distortion driven symmetry breaking, away from the parent tetrahedral (T$_d$) geometry to an ensemble of symmetry-broken covalently bonded structures where all four Cl atoms remain covalently bonded to C. All these bonded forms can be viewed as vibrationally excited C$_{2v}$ \ce{CCl4^+}, as this geometry is predicted by theory to be the only stationary point on the cation potential energy surface with four C-Cl covalent bonds. Theory further indicates that the ions are only temporarily trapped in the bonded form,
based on \textit{ab initio} adiabatic quasiclassical trajectory (QCT) calculations and computed X-ray spectral features from orbital optimized density functional theory (OO-DFT\cite{hait2021orbital}) that can be directly compared to experiment. 
The experiments corroborate the computed spectral features and identify the timescales for the initial symmetry breaking. Similarly, experiment and theory synergistically describe the evolution of the transient covalently bonded cation to a noncovalently bound complex between \ce{CCl3+} and Cl. The latter structure ultimately undergoes irreversible dissociation over relatively long timescales ($\sim$800 fs). 
We also find evidence for a higher energy channel that results in \ce{CCl3} and atomic \ce{Cl^+}, which is expanded upon in the SI.

\section{\label{sec:level1b}Methods}

The tabletop experimental apparatus was previously described in Ref \citenum{Barreau2020}. We provide a brief summary that also touches on relevant upgrades. The pump pulse that induces strong-field ionization is centered at 800 nm, spectrally broadened, and compressed to about 6 fs in duration \textcolor{black}{(as described in the ESI)}. The maximum energy of the compressed pulse at the sample is 150 $\mu$J at 1 kHz repetition rate. It is focused to 65 $\mu$m  full width at half maximum (FWHM) to achieve an electric field with peak intensity up to $3 \pm 1\times 10^{14}$ W/cm$^2$, as estimated by numerical calculations and by observed relative ionization rates of Ar, as explained in the SI.

The X-ray probe is generated by high harmonic generation (HHG), using a 10 fs pulse, centered at 1300 nm, focused into a semi-infinite gas cell filled with 2.3 bar He. This tabletop setup provides X-ray photons with energies up to 370 eV.
The X-ray monochromator has 0.2 eV spectral resolution at the C K-edge and the experimental data are measured as a change in absorbance, or $\Delta$OD, which is separated into spectral components by multivariate fitting, described in detail in the SI. The temporal cross-correlation of the experiment is measured to be $8\pm2$ fs by the autoionization in Ar L$_{2,3}$ lines\cite{Fidler2019FWM} \textcolor{black}{(as discussed in the ESI)}. Experiments are performed over different time ranges to cover a wider range of dynamics; the shortest have a step size of 1 fs and extend to 80 fs, and the longest have variable step sizes and extend to 10 ps. Additionally, experiments are run with varying pump power, from $\sim1-3\times10^{14}$ W/cm$^2$, to assess how the power affects the temporal dynamics and final states. These scans are taken as closely in time as possible to make the comparisons between scans more consistent. Timescales are extracted by fitting lineouts in time to \textcolor{black}{unimolecular kinetics (as described in the ESI)}. Each fit includes convolution with a gaussian of $8\pm2$ fs FWHM to account for the cross-correlation of the experiment, using the 95$\%$ confidence intervals as error bars. \ce{CCl4} was obtained from Sigma-Aldrich at 99.5$\%$ purity and was vaporized by exposing the liquid to vacuum \textcolor{black}{at room temperature. The \ce{CCl4} was probed in a finite gas cell with a 4 mm pathlength with a foreline pressure of 12 mbar.}

Quantum chemical calculations were performed with the Q-Chem 5 software \cite{epifanovsky2021software}. Structures were optimized with the $\omega$B97M-V\cite{wB97MV} density functional and the aug-pcseg-3\cite{jensen2014unifying} basis set. Zero-point energies were found at the same level of theory. Relative ground state electronic energies at the optimized geometries were computed with CCSD(T)\cite{raghavachari1989fifth} extrapolated to the complete basis set (CBS) limit, as detailed in the SI. \textit{Ab initio} adiabatic trajectory calculations on \ce{CCl4^+} were performed with $\omega$B97M-V/aug-pcseg-1, starting from the equilibrium \ce{CCl4} T$_d$ structure and with quasiclassical velocities\cite{karplus1965exchange} for nuclei (this ensures each normal mode of the neutral species has the associated zero point energy). A total of 256 trajectories (out of the 512 possible ones for the \ce{CCl4} molecule with 9 normal modes) were run. The trajectory calculations did not incorporate an external electric field and thus can only provide a first estimate of timescales that can be compared to experiment.

X-ray absorption spectra were simulated with OO-DFT\cite{hait2021orbital}, utilizing the SCAN\cite{SCAN} functional, aug-pcX-2 basis\cite{ambroise2018probing} on the site of core-excitation, and aug-pcseg-2 basis\cite{jensen2014unifying} on all other atoms. This approach has been shown to be accurate to $\sim$0.3 eV root-mean-squared error for the core-level spectra of electronic ground states of both closed-shell\cite{hait2020highly} and open-shell\cite{hait2020accurate} species, without any need for empirical energy translation of the spectra. Refs \citenum{hait2020highly} and \citenum{hait2020accurate} provide detailed protocols for running OO-DFT calculations.  Excited state orbital optimization was done with the square gradient minimization (SGM\cite{hait2020excited}) and initial maximum overlap method (IMOM\cite{barca2018simple}) algorithms, for restricted open-shell and unrestricted calculations, respectively. 

\section{\label{sec:level1c}Results and Discussion}

\subsection{General Features of Experimental Spectrum}
The ground state static spectrum of \ce{CCl4} is presented in Fig. \ref{fig:deltaODFig}a. The Cl L$_{2,3}$-edge comprises a number of peaks, several of which overlap, largely due to the Cl 2\textit{p} spin-orbit splitting of 1.6 eV. The dominant features correspond to excitations to the $7a_1^*$ and $8t_2^*$ symmetry adapted linear combinations (SALCs) of $\sigma_{\ce{CCl}}^*$ antibonding orbitals. Transitions from the L$_3$  (2\textit{p}$_{3/2}$) level to $7a_1^*$ and $8t_2^*$ occur at 200.4 eV and 203.2 eV respectively\cite{lu2009core}. The corresponding L$_2$ (2\textit{p}$_{1/2}$) transitions occur at 202.0 eV and 204.8 eV. All of these excitations are dipole allowed. Transitions to Rydberg levels lead to additional peaks \textcolor{black}{(such as the one at 205.5 eV\cite{lu2009core})}, and a rising ionization edge begins and persists around 207.5 eV. We note that the L$_3$ ionization energy of \ce{CCl4} from X-ray photoelectron spectroscopy is 207 eV\cite{jolly1984core}.  

In contrast, the C K-edge spectrum of neutral \ce{CCl4}  consists of a single intense peak at 290.9 eV\cite{hitchcock1978inner}, followed by a rising edge at $\sim$295 eV. The 290.9 eV peak arises from the C $1s\to 8t_2^*$ excitation, as the transition to the lower energy $7a_1^*$ orbital is dipole forbidden\cite{hitchcock1978inner}. OO-DFT predicts the C$1s\to 8t_2^*$ excitation to be at 290.8 eV, in excellent agreement with experiment, and the forbidden C$1s\to 7a_1^*$ transition is predicted at 289.0 eV. \textcolor{black}{The HHG process leads to lower photon flux at the higher C K-edge energy range than the Cl L-edge, contributing to less signal-to-noise in the former.} There is also significant absorption from the tail of the chlorine edge in the experimental spectrum, which \textcolor{black}{further} reduces the signal-to-noise at the C K-edge. 
 
\begin{figure*}[!t]
\includegraphics[max size={\linewidth}{\textheight/2}]{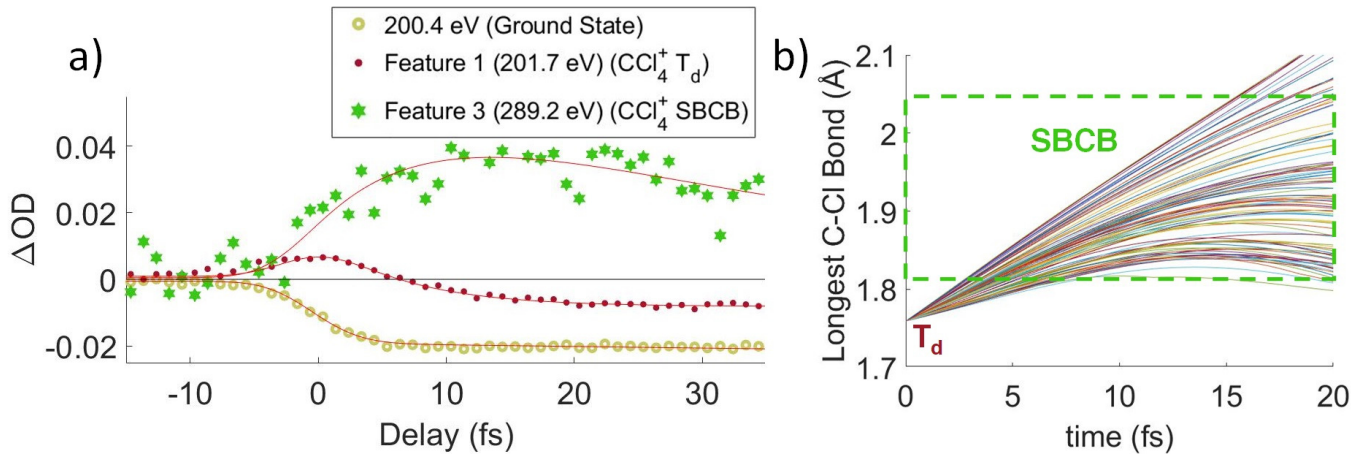}
\caption{\label{fig:trajectoryFig} 
a: Averaged lineouts of features 1 and 3 in the first 35 fs. Fitting including the 8 $\pm$ 2 fs cross-correlation gives a lifetime of 6 $\pm$2 fs and 7 $\pm$ 3 fs for the decay and rise of features 1 and 3, respectively.
b: The longest C-Cl bond length for a set of 100 trajectory calculations, because that bond will correspond to the lowest energy absorption at the C K-edge. It shows that the bond elongates into the symmetry-broken covalently bonded form (labeled SBCB) rapidly, assisting the assignment that feature 1 corresponds to the T$_d$.
}
\end{figure*}

Strong-field ionization by the pump pulse induces dynamics that leads to differences in absorption ($\Delta$OD), as shown in Fig. \ref{fig:deltaODFig}b \& c. The many spectral overlaps at the Cl edge (Fig. \ref{fig:deltaODFig}b) make it difficult to identify individual signals in the $\Delta$OD data. Negative signal from ground state bleach is predominant from 200 to 210 eV with positive signal appearing in the 211-216 eV range, possibly due to a positive local charge on the chlorines leading to a net blue shift. The C edge is easier to resolve, with only a ground state bleach at 290.8 eV, and positive absorption at both higher ($\sim$ 292 eV) and lower (287-290 eV) energies. 

Within these broad experimental signals, several notable features can be observed. The first is a small positive feature at the Cl edge, labeled feature 1 in Fig. \ref{fig:deltaODFig}b. This feature appears transiently at early times and rapidly decays within a few femtoseconds. The second is at 209.5 eV, feature 2, which initially drops to a negative $\Delta$OD but decays back towards zero values on the timescale of 100 fs. A similar timescale is observed in the decay of feature 3, at 289.2 eV, which suggests that the two decays measure the same process. Additionally, the positive signal of feature 3 initially rises on the same time scale as the decay observed in feature 1, suggesting that the early time behavior of features 1 and 3 reflect the same process. A fraction of feature 3 undergoes a continuous evolution (289.2 to 287.1 eV) to feature 4 at 287.1 eV after $\sim$20 fs. Feature 4 continues to grow after that time, concurrent with the decay of feature 3.
Finally, at delays of several hundred femtoseconds to a few picoseconds, much longer times than those shown in Fig. \ref{fig:deltaODFig}, a series of sharp spectral features between 204 and 207 eV become resolvable from the broader features, shown more clearly in Fig. \ref{fig:Cl long time} and described in Sec \ref{sec:clform}. 

Based on the three timescales involved in the evolution of the noted features, few-femtoseconds, tens of femtoseconds, and several hundred femtoseconds, we conclude that three separate processes are responsible for the dynamics of the dissociation of \ce{CCl4+}. 
In order to better understand the nature and quantitative times of these processes, these features are averaged and fitted along the time axis, and assignments are made by comparison to theory, as discussed below. \textcolor{black}{In particular, we note that our probe would not be able to directly track a hole in the Cl lone pair levels of pure 3p character. The Cl 2p$\to$ 3p transition is dipole forbidden, and the C 1s$\to$ Cl 3p process involves local sites on distinct atoms with essentially no overlap. Changes in electronic and nuclear structure arising from such Cl 3p holes would however lead to other transient features, as discussed later. Transitions from Cl 2s levels (L$_1$ edge) to a 3p hole are dipole allowed. However Cl L$_1$ edge transition peaks are very broad and not very intense (as shown through an example in the ESI). We therefore did not pursue analysis in that regime ($\sim$ 270 eV).}

\begin{table}[htb!]
\begin{tabular}{|l|l|l|}
\hline
\begin{tabular}[c]{@{}l@{}}Feature \#\\ Energy (eV)\end{tabular} & Assigned Transition    & \begin{tabular}[c]{@{}l@{}}Time Constant, $\tau$\\ (e$^{-1/\tau\times{}t}$) (fs)\end{tabular} \\ \hline
1 (201.7 eV)       & T$_d$ $\rightarrow$ SBCB & 6 $\pm$ 2                                         \\ \hline
3 (289.2 eV)       & T$_d$ $\rightarrow$ SBCB & 7 $\pm$ 3                                         \\ \hline
2 (209.5 eV)       & SBCB $\rightarrow$ NBC   & 90 $\pm$ 10                                       \\ \hline
3 (289.2 eV)       & SBCB $\rightarrow$ NBC   & 80 $\pm$ 30                                       \\ \hline
4 (287.1 eV)       & NBC Appearance          & \begin{tabular}[c]{@{}l@{}}$\tau$: 50 $\pm$ 20\\ Delay: 23 $\pm$ 8\end{tabular}           \\ \hline
204.2, 206.6 eV      & Atomic Cl Appearance    & 800 $\pm$ 200                                     \\ \hline
214.7 eV      & \ce{Cl+} Appearance     & \begin{tabular}[c]{@{}l@{}}$\tau$: 85 $\pm$ 10\\ Delay: 37 $\pm$ 6\end{tabular}           \\ \hline
\end{tabular}
\caption{\label{table:time constants} Time constants for each of the fits shown in the figures. \textcolor{black}{Further details about the fits are provided in the ESI.}}
\end{table}

\subsection{General Considerations for C K-edge}
 Any distortion away from T$_d$ geometries would alter the symmetry of the $\sigma_{\ce{CCl}}^*$ levels, which would no longer form the $7a_1^*$ and $8t_2^*$ SALCs. The most general symmetry-broken case would be of C$_1$ symmetry, with four unequal bond lengths. This leads to four nondegenerate $\sigma^*$ levels that permit dipole allowed transitions from the C 1\textit{s} level, although intensities will be lower for the MOs with greater C 2\textit{s} character. The shorter bonds would feature stronger C-Cl interactions, leading to higher energy $\sigma^*$ MOs, while the longer bonds would conversely lead to lower energy $\sigma^*$ levels.  In particular, the lowest energy $\sigma_{\ce{CCl}}^*$ level is expected to be dominated by the longest C-Cl bond, and therefore the lowest energy C K-edge feature in the transient absorption spectrum should correspond to this bond. Conversely, the shortest bonds should lead to the highest energy absorption feature. More symmetric configurations can permit multiple bonds to make comparable contributions to any given $\sigma^*$ MO, preventing assignment of a transition to one particular bond. In general, however, the longest bonds should lead to the lower energy features in the C K-edge spectrum, while the higher energy features should arise from shorter bonds. However, there is no such simple rule of thumb available for the Cl L-edges, due to the complexity of the spectrum.

\subsection{\label{}Jahn-Teller Distortion of \ce{CCl4+}}

Lineouts of the few-femtosecond process are shown in Fig. \ref{fig:trajectoryFig}a. These show the decay of feature 1 in $6\pm2$ fs and the rise of feature 3 in $7\pm3$ fs. These lifetimes, along with those extracted from the other fits, are compiled in Table \ref{table:time constants} \textcolor{black}{(with further details about the fits being provided in the ESI)}. The ground state bleach at 200.4 eV from depletion of the neutral \ce{CCl4} represents the instrument response function (i.e., the temporal broadening introduced by the experiment). Comparison of features 1 and 2 to the ground state bleach in Fig. \ref{fig:trajectoryFig}a show these lifetimes are significant beyond the experimental cross-correlation.

\begin{figure*}
\includegraphics[max size={\linewidth}{\textheight/2}]{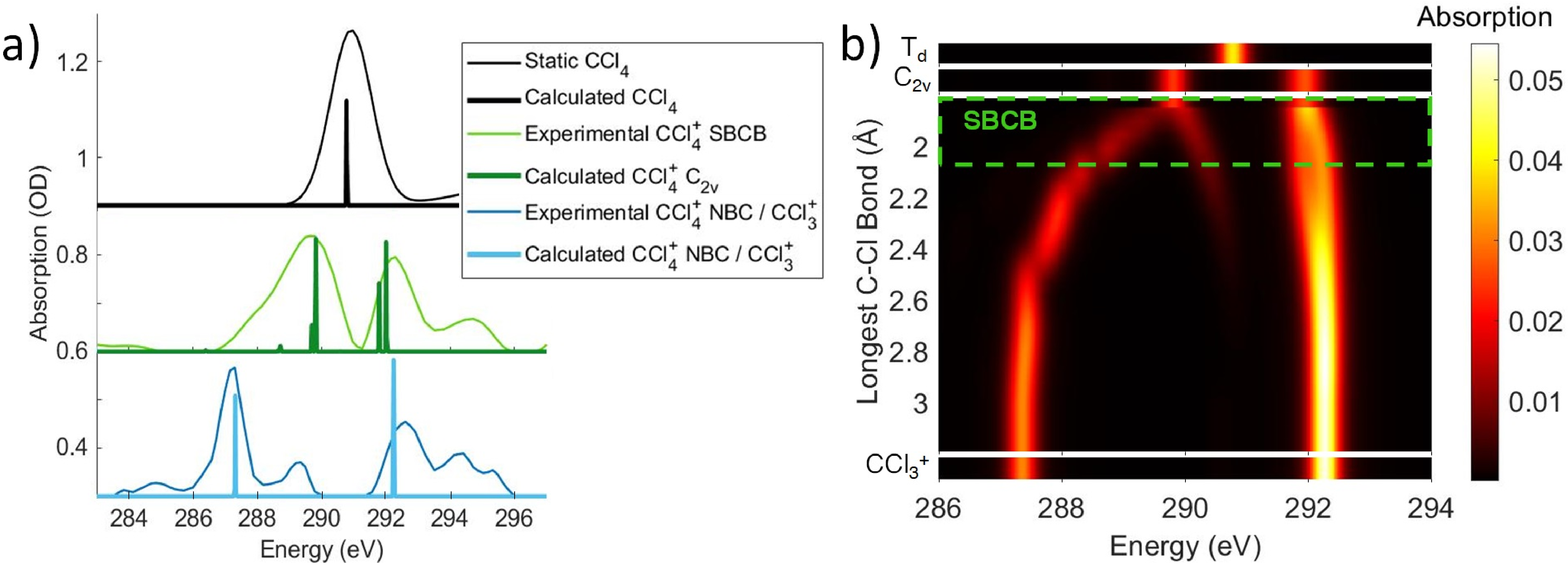}
\caption{\label{fig:spectrumOutput} 
a: The experimental features 3 and 4 as well as the neutral static are disambiguated and corrected for ground state bleach to obtain an absorption spectrum, using a method described in the SI. Notably, this shows that feature 3 $\Delta$OD signal at 289.2 eV corresponds to an actual maximum at 289.8 eV. These spectra are compared to OO-DFT calculations for energies and oscillator strengths of the excitations (given by the peak position and heights, respectively). This allows for assignment of feature 3 to the symmetry-broken covalently bonded form (labled SBCB) and feature 4 to the noncovalent \ce{Cl---CCl3+} complex (labeled NBC) or \ce{CCl3+} moeity.
b: OO-DFT absorption spectra for various \ce{CCl4+} geometries. The third plot from the top shows absorption as a function of a single increasing bond distance (and other nuclear coordinates being optimized with this constraint). This shows splitting of absorption energies with the lower energies corresponding to the longest bond and higher energies to the shortest. The region of maximum bond extension without cleavage is roughly shown in the symmetry-broken covalently bonded region. These longer bonds account for the low energy tail in the covalently bonded experimental spectrum.
}
\end{figure*}

We assume the strong-field ionization process abruptly populates a \ce{CCl4+} cation state or states in a Franck-Condon-like manner, and those states then undergo geometry changes. The states may also have significant internal vibrational excitation. The lowest vertical ionization energy of \ce{CCl4} is 11.7 eV\cite{potts1970photoelectron}, corresponding to loss of an electron from the non-bonding $t_1$ SALC of Cl 3p orbitals and forming the cation X state\cite{potts1970photoelectron,dixon1971photoelectron,Kinugawa2002photoionization}. The electron hole is thus essentially of pure Cl 3p character, and the C $1s\to$ hole transition is consequently of negligible oscillator strength. Indeed, the computed C K-edge XAS of T$_d$ \ce{CCl4+} is quite similar to neutral \ce{CCl4}, due to the extensive delocalization of the hole over all four Cl atoms. None of the features of the  C K-edge experimental spectrum can therefore be unambiguously assigned to T$_d$ \ce{CCl4^+}. At the Cl L$_{2,3}$-edges, the hole density should blue shift the absorption spectrum\cite{Vura-WeisOxidation2013}, and we make a tentative assignment of feature 1 to T$_d$ \ce{CCl4+}.

From theory, the closest local minimum to the T$_d$ starting structure is a C$_{2v}$ symmetry distorted tetrahedron with two long (1.82 {\AA}) and two short (1.70 {\AA}) C-Cl bonds. The calculations indicate that a C$_{2v}$ form of \ce{CCl4^+} is lower in energy by 0.4 eV, and there is no energy barrier between this minimum and the initial T$_d$ geometry. This energy stabilization is smaller than the 1.5 eV stabilization observed for the analogous \ce{CH4} structure\cite{Boyd1991MethaneJTEnergy}. This result is unsurprising as the \ce{CCl4} ionization is from Cl lone-pairs while the \ce{CH4} electron loss is from bonding orbitals. The energy stabilization from JT distortion will be available to the vibrational modes of the ion, so many different nuclear configurations will be accessible around the C$_{2v}$ local minimum configuration. This range of configurations will be referred to as the symmetry-broken covalently bonded form of \ce{CCl4+} (or covalently bonded, for simplicity). 

The OO-DFT C K-edge spectrum of the C$_{2v}$ stationary point was computed in order to determine if the covalently bonded forms were contributing to feature 3. At the C$_{2v}$ stationary point, all four $\sigma^*$ SALCs are nondegenerate and the C 1\textit{s} transitions to these SALCs are all formally dipole allowed. The two lower energy SALCs correspond to the long \ce{C-Cl} bonds, which is computed to lead to absorption at 289.8 eV, and the shorter bonds lead to higher energy $\sigma^*$ SALCs that are computed to absorb at $\sim$292 eV. A comparison of these energies in Fig. \ref{fig:spectrumOutput}a with the absorption of feature 3, corrected for ground state bleach, shows that the energies match well. This validates the assignment that feature 3 arises from distorted \ce{CCl4^+} with four covalent bonds. 

\begin{figure*}
\includegraphics[max size={\linewidth}{\textheight/2}]{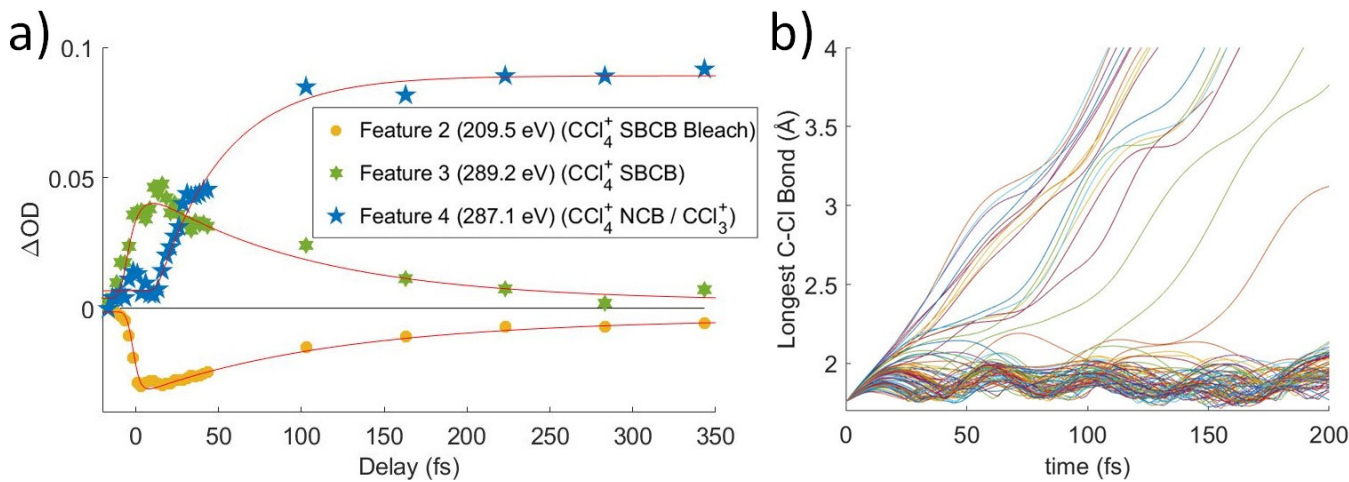}
\caption{\label{fig:lineoutsMedium} 
a: Averaged lineouts of features 2, 3, and 4 out to 350 fs. Fitting gives decay times of features 2 and 3 in 90$\pm$10 fs and 80$\pm$30 fs, respectively. Feature 4 appears only after a delay of 23$\pm$8 fs, followed by \textcolor{black}{a growth signal} with a lifetime of 50$\pm$20 fs. Acronyms: SBCB = symmetry-broken covalently bonded \ce{CCl4+}, NBC = non-covalently bonded \ce{Cl---CCl3+} complex.
b: A set of 100 trajectory calculations, showing the longest C-Cl bond distance. It shows a population that temporarily remains in the covalently bonded form with longest bonds going between 1.8 and 2.05 {\AA}. It also shows a portion of ions have sufficient velocity in the appropriate chlorine to break the C-Cl covalent bond immediately, while others require additional time to redistribute the energy such that the bond may be broken. The trajectories show more population remaining in the covalently bonded form for a longer time, but the difference is likely due to excess vibrational energy in the experiment or another particular of the strong-field ionization process, which are not included in the trajectories. See Sec IX of supporting information for a discussion about the long time behavior of trajectories.
}
\end{figure*}

In order to get an idea of the timescales of this distortion, Fig. \ref{fig:trajectoryFig}b shows the longest C-Cl bond distance for a random subset of 100 calculated trajectories. The longest C-Cl bond is used here as a measure of distortion away from T$_d$, for best comparison to the experimental X-ray absorption. 
It shows that at least one C-Cl bond rapidly elongates to the C$_{2v}$ value of 1.82 {\AA} at $\sim$5 fs on average.

The $\sim$5 fs time from the trajectories is comparable to the 6 $\pm$ 2 fs decay of feature 1 and the 7 $\pm$ 3 fs rise of feature 3. Both time constants should measure the same process, the lifetime of the \ce{CCl4+} T$_d$ geometry, with feature 1 more directly measuring the decay of the initial T$_d$ state and feature 3 measuring formation of the distorted state. They confirm the rapidity of the JT process and the barrierless energy surface between the two. 
We also note that on changing the pump intensity, the lifetimes exhibited no power dependence beyond what is necessary for ionization within the error bounds.

\subsection{\label{}Covalent Bond Breakage of \ce{CCl4+}}
The evolution of some fraction of feature 3 to the much lower energy feature 4 in the experimental spectrum suggests further C-Cl bond stretching. OO-DFT calculations confirm this, with Fig. \ref{fig:spectrumOutput}b showing the absorption of energy in \ce{CCl4^+} as a function of the longest C-Cl bond distance (with all other nuclear coordinates being optimized). Fig. \ref{fig:spectrumOutput}b further explains the lower energy tail of the experimental covalently bonded form spectrum in Fig. \ref{fig:spectrumOutput}a, as the experiment will sample many molecules spanning a wide range of covalently bonded geometries.

Feature 4 at 287.1 eV absorbs at the same energies as an extremely elongated C-Cl distance ($\sim 3$ {\AA}) in Fig. \ref{fig:spectrumOutput}a and b, suggesting that feature 4 corresponds to a \ce{CCl3+} moiety.  \textcolor{black}{Indeed, this feature corresponds to a transition to the unoccupied 2p orbital in \ce{CCl3+}. This \ce{CCl3+} orbital originates from the lowest energy $\sigma^*_{\ce{CCl}}$ level of the covalently bonded form, which has decreasing contribution from the dissociating Cl atom as the C-Cl distance increases.
The observed spectrum} is thus consistent with the theoretical prediction that the lowest energy forms of \ce{CCl4+} are noncovalently bound complexes between atomic Cl and \ce{CCl3+}. The global energy minimum of \ce{CCl4+} is a complex with approximately C$_{3v}$ symmetry, a nearly planar \ce{CCl3^+} moiety having a Cl atom vertically above the C, at a distance of 3.4 {\AA}. A C$_s$ symmetry minimum with the atomic Cl coordinating to a bonded Cl in the \ce{CCl3+} moiety is also found at 0.02 eV above the minimum energy complex, with a Cl-Cl distance of 3.24 {\AA}. 
Given the only slight energy preference for the minimum energy position of the Cl in the complex 
and the available vibrational energy, it is likely that the Cl does not stay at this position under experimental conditions and instead samples a wide range of locations around \ce{CCl3^+}. However, the experiment is not directly sensitive to the position of the noncovalently bound Cl. The large distances between \ce{CCl3^+} and Cl for all such species in fact indicate that C K-edge transitions to valence orbitals would be essentially unaffected by the particulars of the noncovalent interaction. 

\begin{figure*}
\includegraphics[max size={\linewidth}{\textheight/2}]{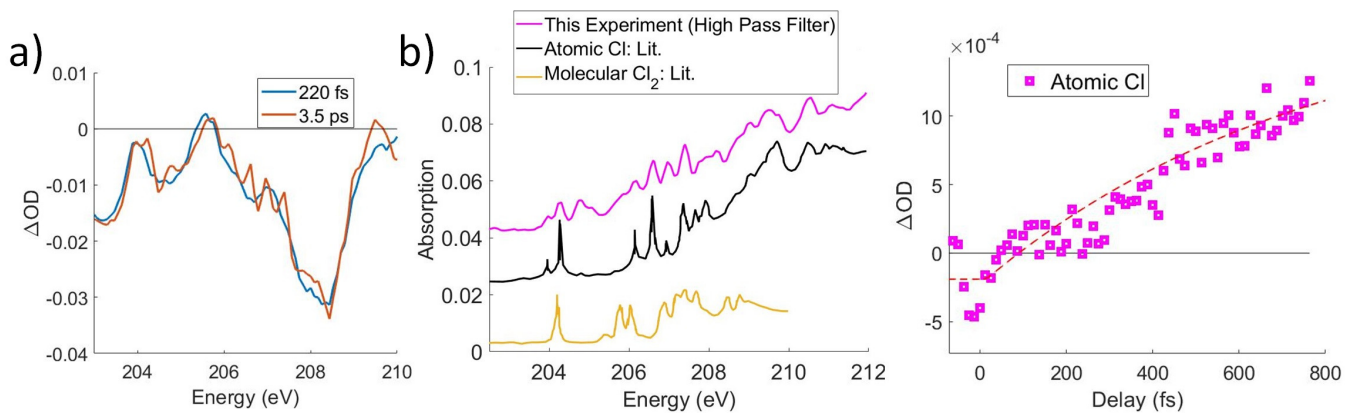}
\caption{\label{fig:Cl long time} 
a: Lineouts in energy for 220 fs and 2.5 ps at the Cl L$_{2,3}$-edges. They show that although the overall shape and intensity of the $\Delta$OD signal remains constant, sharp spectral features appear at longer delays. The difference between 3.5 ps and 220 fs is taken and added to an error function to obtain the spectrum in b.
b: A comparison of the sharp spectral features is made to atomic Cl\cite{Caldwell1999AtomicCl} and to molecular \ce{Cl2}\cite{Nayandin2001Cl2}, which shows that the features match well with atomic Cl.
c: A high-pass spectral filter is applied to the $\Delta$OD data and an average of the two most prominent lines at 204.2 and 206.6 eV are a lineout is shown against time. A fit is taken starting from t=0, due to noise preventing a better determination of a start point. This fit with $\sim3\times10^{14}$ W/cm$^2$ pump intensity has a lifetime of 800 $\pm$ 200 fs.
}
\end{figure*}

An energy decomposition analysis (EDA\cite{mao2021intermolecular}) calculation reveals that the -0.15 eV interaction energy between atomic Cl and \ce{CCl3^+} in the minimum energy complex geometry is mostly (56\%) from the polarization of the atom by the cation, with a much smaller amount (19\%) arising from charge-transfer (the remaining 25\% being permanent electrostatics, Pauli repulsion and dispersion). The system effectively acts as a charge-induced dipole complex.  It is also worth noting that these noncovalent complex structures do not have any covalent Cl-Cl interactions, contrary to some early assignments\cite{DrewelloCCl4plus,KimeCCl4plus,MutoCCl4plusSolidESR}.

The timescale of the symmetry-broken covalently bonded \ce{CCl4+} species to \ce{Cl---CCl3+} complex transition is shown through lineouts of features 2, 3, and 4 in Fig. \ref{fig:lineoutsMedium}a. This shows decay times of features 2 and 3 in 90$\pm$10 fs and 80$\pm$30 fs, respectively. Feature 4 appears only after a delay of 23$\pm$8 fs, which implies that the nuclear motions take at least 23 fs to get to the point that the ion starts to resemble its dissociated form. After this delay, it has an additional lifetime of 50$\pm$20 fs towards reaching its asymptotic value. Because of the delay from nuclear movement, the experimental time for the covalently bonded to noncovalent complex process is taken from the more precise feature 2, 90$\pm$10 fs, and no pump power dependence is observed. This time is expected to be significantly longer than the JT distortion due to the computed barrier of 0.17 eV for this process. In varying the pump power, we do not observe a difference in the covalently bonded population decay times, possibly due to the large intensity already required to ionize the \ce{CCl4}.

The trajectory calculations in Fig. \ref{fig:lineoutsMedium}b show that a portion of the ions continue their initial distortion and dissociate immediately, which is observed experimentally by the direct evolution of the feature 3 $\to$ feature 4 signal. \textcolor{black}{Indeed, this regime of the experimental signal reports direct dynamics from the covalently bonded form to the complex for a fraction of the molecules.} Another portion remains trapped in the covalenly bonded form for longer than a vibrational period, showing longest C-Cl distances between 1.8 and 2.05 {\AA}, which serves to define the symmetry-broken covalently bonded region. We however note that the trajectory calculations show a significant population persisting in the covalently bonded form for a longer time than the experimentally measured lifetime. This likely arises from the trajectory calculations not including the excess vibrational energy added by the pump pulse. The calculations only include the zero-point energy of neutral \ce{CCl4}, leading to the covalently bonded form persisting longer than what is experimentally observed.  A further discussion about the long time behavior of the trajectories is given in Sec IX of the ESI.

\subsection{\label{sec:clform}Atomic Cl Dissociation}

Although the C K-edge cannot distinguish between noncovalent complex and free Cl, the Cl L$_{2,3}$ edge can make this distinction, as atomically sharp Rydberg lines corresponding to atomic Cl become evident upon completion of the dissociation, which is shown in Fig. \ref{fig:Cl long time}a and b. Very little change is observed in the $\Delta$OD spectra between 250 fs and 3.5 ps at either the C K-edge or chlorine L$_{2,3}$-edge, other than the appearance of these sharp lines, which is consistent with the previous assignment of the noncovalent complex. While weakly bound in the complex, the atomic lines are broadened by a combination of two factors. Firstly, the transitions are already of low intensity due to their Rydberg character, the lowest energy lines being of $2p\to4s$ character. Secondly, the diffuse nature of the Rydberg levels means that the excitation energy would be sensitive to the precise position of the \ce{CCl3+} entity, and the flatness of the complex ground state potential energy surface permits a large range of accessible nuclear configurations. 
However, over time, the complex irreversibly dissociates, leading to atomic Cl lines at 204.2 eV and upward from 206 eV, which agrees with previous experimental absorption data\cite{Caldwell1999AtomicCl}. Thus, we can use the Cl atomic lines to track the time it takes for the complex species to completely dissociate to CCl$_3^+$ and Cl. We note that molecular Cl$_2$ has absorption in this energy range as well\cite{Nayandin2001Cl2}; however, comparison to experimental atomic absorbance in Fig. \ref{fig:Cl long time}b shows much better agreement with atomic Cl. 

\textcolor{black}{A high-pass filter is applied to the transient spectrum to separate the sharp atomic Cl lines from broader molecular signals.} A simple exponential fit of these lines in Fig. \ref{fig:Cl long time}c gives a lifetime longer than 800 fs. The exponential starts at time zero, despite the Cl signal not clearly departing from 0 until $\sim$300 fs. This may correspond to the minimum time required for atomic Cl to move far enough to show the Rydberg lines, although the signal-to-noise of this experiment is not sufficient to make that distinction. 

The 800 fs time scale is much longer than the 90 fs observed for noncovalent complex formation, despite the excess of energy available to the system. However, the interactions and thus rate of energy transfer between the \ce{CCl3+} moeity and the neutral Cl are much weaker than a covalent bond. This situation is similar to van der Waals complexes, which sometimes show lifetimes exceeding milliseconds with vibrational energy in the molecular moeity\cite{nesbitt2012toward,hernandez2012theoretical}. The 800 fs value is the fastest Cl formation time and was measured from the most intense pump pulse of $3\times10^{14}$ W/cm$^{2}$. Datasets collected at lower pump power show slower dissociation times, and some do not exhibit significant atomic Cl formation up to 3 ps. The noncovalent complex signal from feature 4 at 287.1 eV is still present, suggesting that the strong-field ionization may be capable of forming long-lived noncovalently bound \ce{CCl4+} complexes; although times longer than a few ps are outside the scope of this paper. It has been observed that otherwise unstable parent ions can be generated by few-cycle strong field ionizing laser pulses, such as tetramethyl silane and \ce{CS2}\cite{dota2012intense,mathur2013carrier}, and an analogous complex may be a potential explanation of these signals. A quantitative comparison of the power dependence on this timescale was not carried out. \textcolor{black}{We also note that the lifetime of a molecular species against collision with an electron (released by strong-field ionization) is on the order of $\sim 5$ ps with our setup (as discussed in the ESI), which is an additional factor to consider, but only at very long times.}

\section{\label{sec:level2}The High Intensity Case and the Formation of C\MakeLowercase{l}$^+$}

While the observed dissociation to CCl$_3^+$ and Cl can be well explained by the intermediate forms and pathways discussed above, another channel appears to be present in the data. This channel is clearest at the chlorine L$_{2,3}$-edge where very sharp lines with similar widths to the atomic Cl lines are observed at 214 eV with less obvious components at 212.6 eV, shown in the SI. These energies are in the range that the calculations predict for the Rydberg states ($2p\to 3d,4s$) of Cl$^+$. These atomically sharp peaks begin to appear with a delay of $37\pm6$ fs relative to the onset of the main cationic CCl$_4$ signal, and from that point, they show a time to grow in of $85\pm10$ fs for the best signal-to-noise dataset. 

Low energy satellite features in the C 1\textit{s} spectrum $\sim285.5$eV are also observed, which can correspond to transitions to the singly occupied level of the \ce{CCl3} radical, computed to be at 285.5 eV. These simultaneous features appear to suggest the existence of a channel that results in CCl$_3$ and Cl$^+$. The formation of Cl$^+$ is a much higher energy channel, with final energies about 4.8 eV higher than the normal dissociation channel. Cl$^+$ formation has been observed in previous \ce{CCl4} ionization experiments in low quantities from electron impact\cite{Lindsay2004ElectronImpact}, single-photon ionization\cite{Kinugawa2002photoionization,Burton1993Photoionization400eV} and in higher quantities from strong-field ionization\cite{geissler2007concerted}. Further discussion about \ce{Cl+} formation is provided in the SI.

\section{\label{sec:level3}Conclusions}

In this paper, we have shown experimental and theoretical evidence for both a transient Jahn-Teller distorted symmetry-broken covalently bonded \ce{CCl4+}, similar to the stable ion of \ce{CH4+}, and a noncovalently bound complex between \ce{CCl3+} and \ce{Cl}. Neither of these intermediates have been observed previously in experiments. A summary of each of the time constants experimentally extracted is shown in Table \ref{table:time constants}. The transition from the initial tetrahedral ion to the C$_{2v}$ state occurs in 6$\pm$2 fs, showing that the Jahn-Teller distortion is very fast, on the order of the duration of the pump laser pulse. 
It provides an order of magnitude estimate for other symmetric molecules undergoing JT distortion, which can aid chemical simulations as well as help predict the vibrational energies and nuclear dynamics of those other systems, especially given that this symmetry-broken covalently bonded intermediate was not assumed to exist prior to this work\cite{Tachikawa1997CCl4Dissociation, harvey2014photoionization}.

The noncovalent \ce{Cl---CCl3+} complex is bound largely by the polarization of the neutral species, forming an ion-induced dipole complex with a lifetime much longer than the covalent bond cleavage time. The final dissociation time for free Cl or \ce{Cl+} seems to be slightly dependent on the strong-field power, but for the highest power used in this experiment, the lifetime is more than 800 fs (vs 90 fs for noncovalent complex formation). This observed long lifetime is a potential explanation for otherwise unstable parent ions appearing in mass-spectra of ultra-short strong-field ionization.



\section*{Acknowledgements}
This work is funded by the DOE Office of Science, Basic Energy Science (BES) Program, Chemical Sciences, Geosciences and Biosciences Division under Contract no. DE‑AC02‑05CH11231, through the  Gas Phase Chemical Physics program (A.D.R, V.S, E.A.H, E.R, D.M.N, and S.R.L.) and Atomic, Molecular, and Optical Sciences program (D.H. and M.H.G.). The instrument was built with funds from the National Science Foundation through NSF MRI 1624322 and matching funds from the Lawrence Berkeley National Laboratory, the College of Chemistry, the Department of Physics, and the Vice Chancellor for Research at UC Berkeley.  A.D.R. is additionally funded by the U.S. Department of Energy, Office of Science, Office of Basic Energy Sciences, Materials Sciences and Engineering Division, under Contract No. DE-AC02-05-CH11231 within the Physical Chemistry of Inorganic Nanostructures Program (KC3103) and by the W.M. Keck Foundation Grant No. 042982. M.B.B. was funded through NSF (REU Site: Engineering Applications of Extreme Ultra-Violet (EUV) Laser Light) Award No. 1852537. 

\section*{Author contributions statement}
A.D.R. and D.H. contributed equally to this work. A.D.R, V.S, E.A.H, E.R, and M.B.B. performed experiments. A.D.R. analyzed experimental data. D.H. performed calculations. S.R.L, M.H.G, and D.M.N. supervised the project. A.D.R. and D.H. wrote the manuscript, with inputs from all the authors. All authors reviewed the manuscript. 

\section*{Supporting information}
PDF: More details about experimental protocols, fits, computational details, computed spectral features, discussion about \ce{Cl+} formation, changes with varying pump intensity.\\
ZIP: Computed geometries of species.\\
XLXS: Relative energies of species. 
\section*{Competing interests} 
The authors declare the following competing interest: M.H.-G. is a part-owner of Q-Chem, which is the software platform in which the quantum chemical calculations were carried out. 

\onecolumngrid
\appendix


\section{Treatment of Experimental Data}
The experimental data was calibrated from CCD pixel to energy by comparison to known energies in \ce{SF6}\cite{Hudson1993SF6Absorption}, Ar\cite{chew2018ArAbsorption}, and \ce{CO2}\cite{Eustatiu2000CO2Absorption}. Transient data was collected in pump-on/pump-off configuration to collect a change in absorption, $\Delta$OD:
\begin{align}
\Delta{}\text{OD} = \left(-\log_{10}\left(\dfrac{I_{on}}{I_0}\right)\right) - \left(-\log_{10}\left(\dfrac{I_{off}}{I_0}\right)\right) = -\log_{10}\left(\dfrac{I_{on}}{I_{off}}\right)
\end{align}
where $I_{on}$ is the spectrum measured with the pump on and $I_{off}$ is the spectrum with pump off. The reference or unabsorbed spectrum, $I_0$, is eliminated through use of the logarithms and is \textcolor{black}{therefore not collected at each iteration}. The data was filtered using edge reference filtering\cite{Geneaux2021XUVNoise}. Separate edge referencing was done for the chlorine L$_{2,3}$-edges and the carbon K-edge.

\color{black}
\section{Method of Fitting Lineouts to Unimolecular Kinetics}
The key transformation being studied in this work (as depicted in Fig. 1 of the main manuscript) is the following three step process
\begin{align}
    \text{T}_d \ce{->[k$_1$]} \text{SBCB} \ce{->[k$_2$]} \text{NBC}\ce{->[k$_3$]}\text{\ce{CCl3+}+Cl}
\end{align}
where the symmetry-broken covalently bonded form is abbreviated as SBCB, and the non-covalently bonded \ce{Cl---CCl3+} complex is abbreviated as NBC. These processes are assumed to be unimolecular (due to low gas pressure, as discussed later in Sec \ref{sec:collision}) and irreversible. They 
have associated lifetimes $\tau_1=\dfrac{1}{k_1},\tau_2=\dfrac{1}{k_2},$ and $\tau_3=\dfrac{1}{k_3}$, which were reported to be 6$\pm$2 fs, 90$\pm$10 fs and $>800$ fs in the main manuscript. Herein, we discuss how these values were obtained via fitting to lineouts. 

It superficially seems natural to directly fit to the sequential, three-step reaction. However, we instead separate these processes into two stages
\begin{align}
    \text{T}_d \ce{->[k$_1$]} \text{SBCB} \ce{->[k$_2$]} \text{NBC}
\end{align}
and
\begin{align}
   \text{NBC}\ce{->[k$_3$]}\text{\ce{CCl3+}+Cl}
\end{align}
They are assumed to be effectively decoupled from each other. The rationale for this separation is described at the end of this section. 

The first stage can be modeled by rate equations: 
\begin{align}
\dfrac{d[\text{T}_d]}{dt} &= -k_1[\text{T}_d]\\
\dfrac{d[\text{SBCB}]}{dt} &= k_1[\text{T}_d] - k_2[\text{SBCB}]\\
\dfrac{d[\text{NBC}]}{dt} &= k_2[\text{SBCB}]
\end{align}
in terms of populations $[\text{T}_d],[\text{SBCB}]$, and $[\text{NBC}]$.
Integrating the above differential equations\cite{kineticsBook2019}, we find that the populations are given by:
\begin{align}
    [\text{T}_d] &= [\text{T}_d]_0 \times e^{-k_1t}\label{eqn:tdform}\\
    [\text{SBCB}] &= \dfrac{k_1[\text{T}_d]_0}{k_2-k_1} \times (e^{-k_1t} - e^{-k_2t})\label{eqn:sbcbform}\\
    [\text{NBC}] &= [\text{T}_d]_0  - [\text{T}_d] - [\text{SBCB}]\label{eqn:nbcform}
\end{align}

For fitting the lineouts, we fit the absorbances of the T$_d$, SBCB, and NBC at the energy of the lineout, as well as the rates $k_1$ and $k_2$, represented as the lifetimes $\tau_1=\dfrac{1}{k_1}$ and $\tau_2=\dfrac{1}{k_2}$.

This model makes the following assumptions:
\begin{enumerate}
    \item The transition between each of the ion forms are unimolecular and irreversible, which is likely, given that only a single molecule is involved and all steps are energetically favorable.
    \item No other species or processes are involved in significant amounts (as discussed later). 
    \item \textcolor{black}{Fitting based on the differential equations assumes that individual molecules transition between two configurations at a rate that is much faster than the rate of the reaction. In other words, the lifetime of any intermediate form (along the reaction path) is much smaller than the reported lifetimes. 
    This is rather unlikely to be the case for the processes in the first few tens of femtoseconds. However, the model can still be used as a measure of how similar the average population of the molecules are to each of the two limiting configurations (i.e. `reactant' and `product') if the change in absorption for any intermediate form can be found via linear interpolation between the limiting configurations, at a given energy lineout. A case where this assumption would not apply is the absorption of the NBC at 287 eV. Zero absorption is observed therein until $\sim$23 fs, due to the fact that no absorption can occur until a C-Cl bond distance of $\sim$2.5 {\AA} is reached. For this reason, lineouts are chosen at energies where absorption is present for all species of interest. }
\end{enumerate}
The fitting also assumes that $\tau_1 < \tau_2$ even though the shape of the intermediate population $[\text{SBCB}]$ would be similar in the opposite case. This assumption is made, based on the theoretical prediction that the Jahn-Teller distortion to SBCB should be a barrierless process, where the bond cleavage to NBC should have a barrier of 0.2 eV. Additionally, the data shows the shift of absorbance from 289.6 eV to 287.1 eV starting around 14 fs, which would only be expected if the Jahn-Teller distortion to SBCB were completed relatively quickly.

Our fits generally agree well with the data, which shows that the exponential nature of the fits is sufficient, within the signal-to-noise of the experiment. 



The $\text{NBC}\ce{->[k$_3$]}\text{\ce{CCl3+}+Cl}$ process is more difficult to fit, as the NBC and \ce{CCl3+} have rather similar absorption profiles, especially at the energies chosen for fitting $k_1$ and $k_2$. The signals associated with the NBC dissociation to form free Cl are very weak when integrated across many energies. However, fitting can be done via applying a high pass filter, which eliminates the broad molecular peaks while leaving behind atomic Cl signals. The growth in these signals (at 204.2 eV and 206.6 eV) is fit to:
\begin{align}
     [\text{Cl}] &= [\text{Cl}]_\infty\times (1- e^{-k_3t})
\end{align}

The decoupling of this last step from preceding steps can be justified by noting that the resulting $\tau_3=\dfrac{1}{k_3}\sim 800$ fs is considerably longer than the previously obtained $\tau_1,\tau_2$. Therefore, Eqns \ref{eqn:tdform} and \ref{eqn:sbcbform} indicate $<10^{-4}$ fraction of the molecules are left in either the T$_d$ or SBCB configurations. Thus, NBC formation is effectively complete on the timescale of Cl dissociation. Additionally, the energies of the lineouts for the $k_1$ and $k_2$ kinetics were chosen such that the change in absorption to the last step is zero, and the high-pass filter for the atomic Cl eliminates the effects of the first two processes. These processes can therefore be neglected while modeling the last step. 

\color{black}


\textcolor{black}{\section{Pump Pulse and Cross-Correlation}}

\FloatBarrier
\begin{figure}[htb!]
\includegraphics[max size={\linewidth}{\textheight/2}]{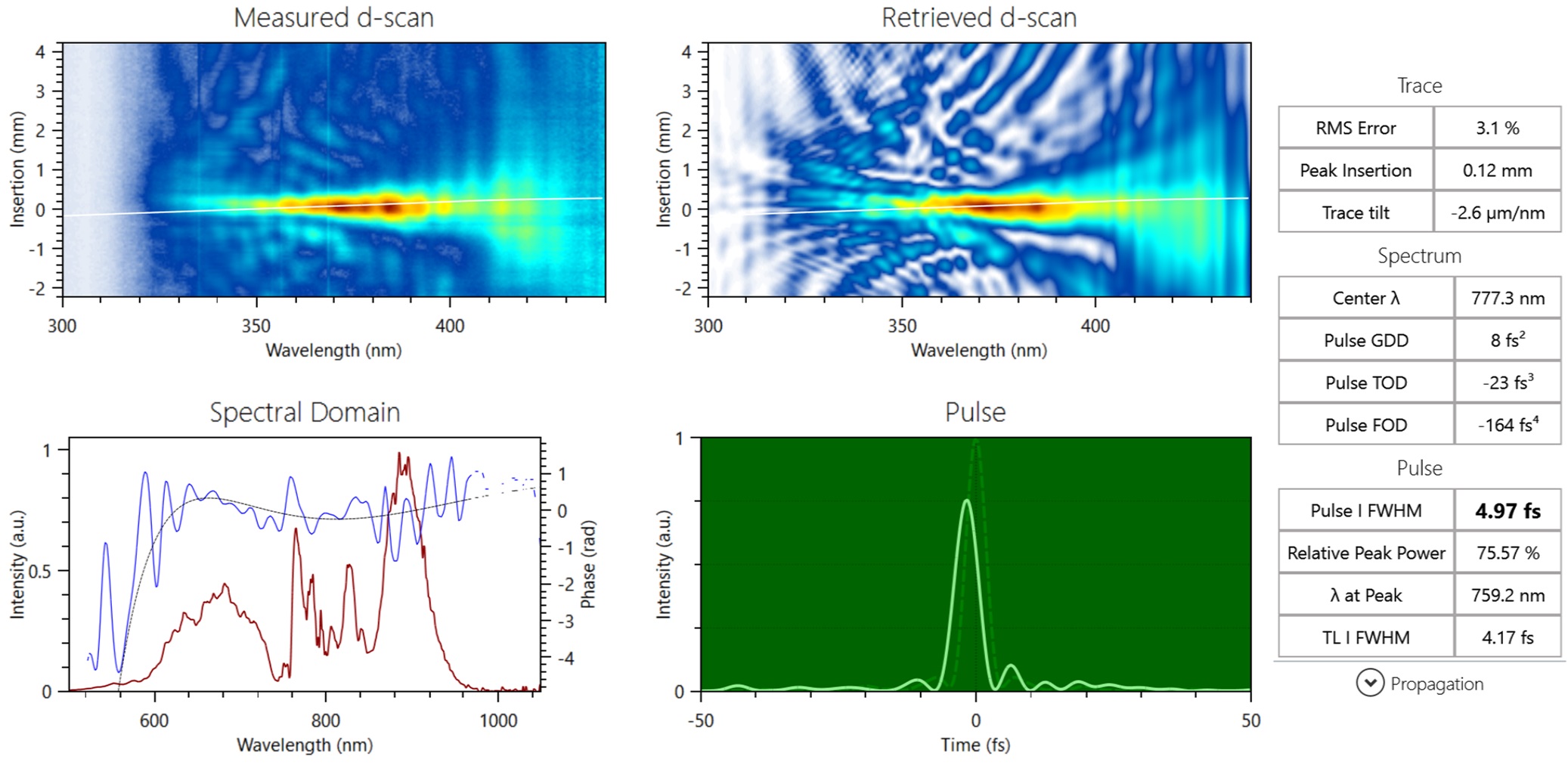}
\caption{\label{fig:dscanData} 
\textcolor{black}{Output of the d-scan software\cite{miranda2013ultrashort,silva2014simultaneous}. Top-left: Measured, calibrated SHG d-scan as a function of fused-silica insertion. Top-right: Numerical retrieval of measured d-scan. Bottom-right: Measured input spectrum (red) and retrieved spectral phase (blue). 4$^{th}$ order polynomial fit of the phase (black). Bottom-right: Retrieved temporal intensity profile of the measured pulse (light-green) and temporal intensity of the transform limited pulse (dashed green).}
}
\end{figure}

\textcolor{black}{The pump pulse was characterized by a Sphere Ultrafast Photonics d-scan, which retrieves the short pulse duration and the temporal structure of the pulse\cite{miranda2013ultrashort,silva2014simultaneous}. The measurements were taken outside of the vacuum chambers with compensation glass and air to mimic the path into the vacuum chambers. Over the many days of measurements, the shortest duration was found to be 3.75 fs and the longest was 6 fs FWHM. The latter is reported in the main text, to be conservative. A representative retrieval is shown in Fig. \ref{fig:dscanData}, which shows a 5 fs pulse. No long temporal tails or significant pre or post pulse effects are observed. The pump pulse is linearly polarized.}

\begin{figure}[htb!]
\includegraphics[max size={\linewidth}{\textheight/2}]{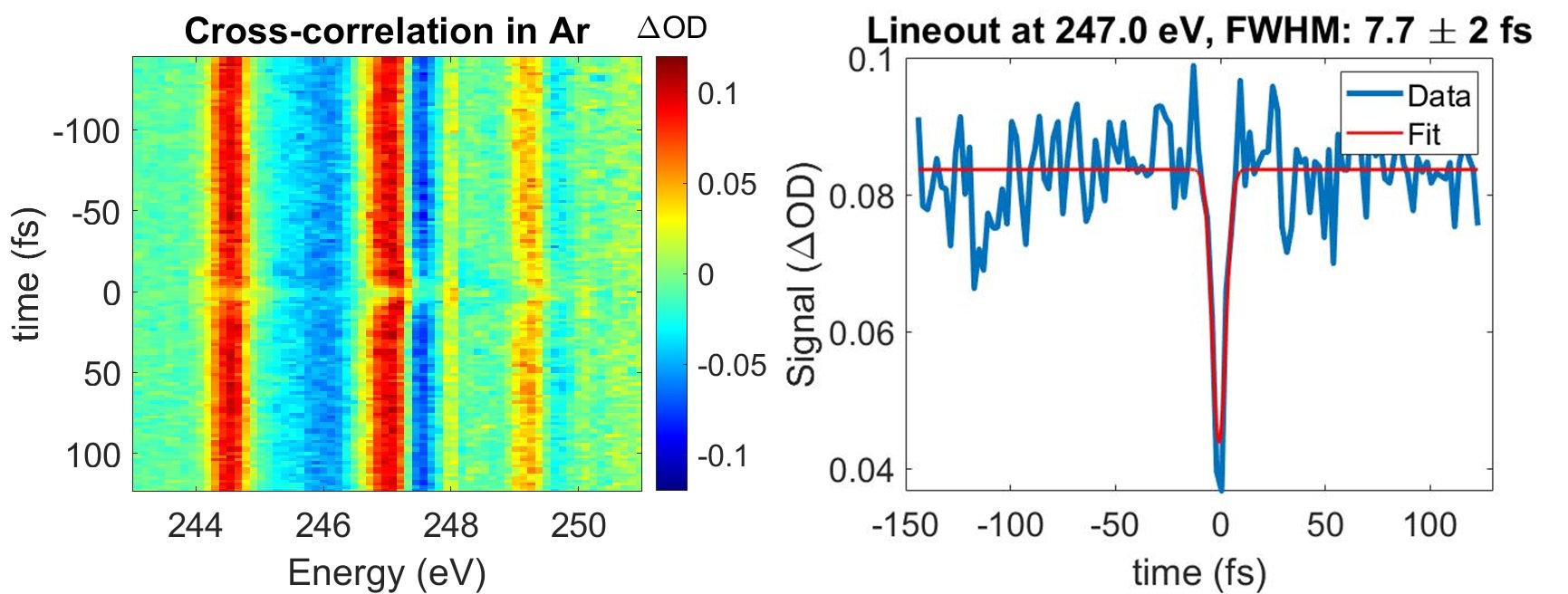}
\caption{\label{fig:ArCrossCorrelation} 
\textcolor{black}{Cross-correlation as determined by suppression of Ar L$_{2,3}$ lines. Left: Transient $\Delta$OD plot, using a high-pass filter\cite{ottNature2014}, the positive (red) are the sharp features of the unperturbed absorption. The feature at time zero corresponds to interaction with the pump pulse. Right: A lineout is taken at the $2p^{-1}_{3/2}3d$ line (247.0 eV) and fit to a Gaussian, resulting in a fit of 8 $\pm$ 2 fs cross-correlation. }
}
\end{figure}

\textcolor{black}{The cross-correlation of the experiment was determined by suppression of atomic absorption lines in Ar by perturbation of autoionization\cite{chew2018ArAbsorption,cao2016attosecond}. The measurement was taken as transient absorption with 1 fs steps, $\sim1\times{}10^{13}$ W/cm$^2$ pump intensity, and 1000 ms camera integration. The data was filtered by the method described by Ott et.al. to increase the signal-to-noise ratio\cite{ottNature2014}. The suppression of the $2p^{-1}_{3/2}3d$ line was fitted to a Gaussian and the FWHM was found to be 8 $\pm$ 2 fs, as shown in Fig. \ref{fig:ArCrossCorrelation}. This cross-correlation covers many broadening factors, such as duration of pump and probe pulses, timing jitter between pump and probe pulses, and unlocked carrier-envelope phase. These factors will lead to a Gaussian broadening of the signal in the time domain and are included in all kinetic fits in the paper.
}

\section{On the Presence of CC\MakeLowercase{l}$_4^+$ Excited States }

In the main text, comparisons to the simulated spectra are done for the ground (X) state of \ce{CCl4+}. A potential concern is whether excited states of \ce{CCl4+} could provide an alternative explanation or whether such states are present at all. \textcolor{black}{During the strong-field ionization, the large amount of available energy could potentially lead to excited states of the cation. A simple picture of strong-field ionization assumes that it leads to ionization into the ground state of the cation\cite{YuSFIReview,ibrahim2014tabletop}; however, excited states are possible through mechanisms such as recollision of the electron due to the electric field\cite{hao2022recollision,haan2007recollision}. Other studies have shown that ionization from lower electronic levels can be enhanced, based on the orientation of the molecule, relative to the polarization of the electric field\cite{smirnova2009strong,li2020attosecond}. Additionally, there are other pump schema used in other literature that can reduce or disambiguate the effects of such excited states. These pump schema can include: scanning the angle of the polarization or changing the ellipticity of the polarization\cite{lai2016polarization}, modification of the pump pulse duration or spectral phase\cite{mathur2013carrier,Levis2002SFInewproducts}, changes in wavelength of the pump pulse\cite{wolter2015strong}, or use of pre-pulses to orient the molecule or prepare it in a vibration or electronic state\cite{wagner2006monitoring,song2017orientation}. This paper only uses a linearly polarized pump laser and measures the differences in intensity, so the presence or lack thereof of excited states must be infered from $\Delta$OD data.} It is difficult to completely rule out such states, but they likely do not contribute significantly to the \ce{CCl3+} + \ce{Cl} channel. They do however likely lead to \ce{Cl+} formation \textcolor{black}{ as a side channel}. \textcolor{black}{It is also possible that the molecule is ionized to such higher energy excited states and quickly relaxes to the ground state\cite{Remacle2021CH4JT}. However, our method is not able to distinguish these possibilities, and it would not strongly affect the mechanisms described.}

We consider \textcolor{black}{four} classes of excited states, \textcolor{black}{and analyze their potential relevance via insight from quantum chemical calculations}:

\begin{enumerate}
    \item \textbf{Ionization from a deeper level:}  The A and B states of the cation are relatively readily accessible\cite{harvey2014photoionization}, being ionizations from other 3p lone pair SALCS ($t_2$ and $e$) at 12.5 and 13.5 eV respectively. For each of these, there would be no perceptible difference in the Franck-Condon absorption at the C K-edge, because the electron is coming from purely Cl orbitals. Optimization of the A state starting from the initial T$_d$ geometry leads to no local minima, but goes straight to the NBC form (unlike the X state, which leads to the SBCB form). We observe a delay in the formation of the 287.1 eV NBC feature, which would indicate that the bond is not dissociating immediately, but going through the SBCB forms. 
    
    The B state optimizes to a D$_{2d}$ minima (all four bonds of equal length, but angles deviating from T$_d$) and indeed experimental vibrational spectrum for this form has been observed in \textcolor{black}{photoionization experiments}\cite{harvey2014photoionization}, \textcolor{black}{indicating it can exist in a metastable form}. However, this  D$_{2d}$ geometry B state is computed to absorb very close to the static spectrum of neutral \ce{CCl4} at the C K-edge and is thus invisible to that probe.

    The C ($\sim$16 eV) and D ($\sim$20 eV) states are ionizations from the $t_2$ and $a_1$ bonding SALCs ($\sigma_{\ce{CCl}}$ bonding orbitals), respectively, and thus would have C K-edge excitations at $\sim$278-282 eV. Only the C state should be bright, as the transition to the D state hole is dipole forbidden at the C K-edge. Both the C and D states would also strongly absorb in the Cl L-edge at ~187-190 eV, which is consistently zero in the $\Delta$OD data. The absence of any features there suggest lack of holes in the bonding levels in general. 
    
    Beyond the C and D states, ionization from the Cl 3s lone pairs is a possibility, albeit at rather high ($\sim$ 27 eV) energies. This should lead to bright absorption at low Cl L edge energies $\sim$ 180 eV that do not appear in the experiment, although such states would be dark to the C K-edge probe. Subsequent ionizations would occur from the L-edge itself ($\sim$ 207 eV) and are unfeasible with the pump used.
    
    \item \textbf{States where an excitation from a doubly occupied level occurs simultaneously with ionization}: This leads to states with 3 unpaired electrons, \textcolor{black}{a class} of excited states that would not be possible from photoionization, so does not have widely accepted \textcolor{black}{nomenclature}. \textcolor{black}{The lowest energy states of this nature will be excitations from another Cl 3p lone pair level. Our probe cannot directly detect excitations into Cl 3p holes, as discussed in the main text. Therefore, signals from such states will not be immediately apparent. Indeed, such} 
    states could possibly lead to \ce{Cl+}, as noted later. One key observation is that these states cannot directly evolve to ground state \ce{CCl3+}+\ce{Cl}, as the latter has only one unpaired electron. A state crossing that pairs two of the unpaired electrons could permit this, but preliminary computational investigations failed to reveal any such crossing. \textcolor{black}{An alternative possibility is formation of excited state \ce{CCl3+} or \ce{Cl}. The observed atomic Cl signals agree well with previous experimental results for ground state atomic \ce{Cl}\cite{Caldwell1999AtomicCl}. The S$_1$ and T$_1$ excited states of \ce{CCl3+} appear to absorb strongly at 285.7 eV from the C 1s $\to$ C 2p transition. No significant signal in that regime is however observed at the C K-edge, relative to the prominent ground state \ce{CCl3+} signal at 287.1 eV. Excited state \ce{CCl3+} formation therefore appears to not be particularly relevant.}
    
    \textcolor{black}{We also note that excitation of an electron from the bonding $t_2$ or $a_1$ levels (instead of the higher energy Cl 3p lone pairs) would lead to bright low energy spectral features similar to the ones predicted for the C/D states of the cation (from the core $\to$ hole transitions). Similar behavior can be expected for an excitation from the Cl 3s lone pair levels as well. Absence of such signals indicates the lack of such states.}

    \item \textbf{The unpaired electron of the ion is excited to a higher energy level:} This should be higher in energy than the previous type \textcolor{black}{(for an excitation from a doubly occupied 3p lone pair)} and might be expected to also lead to \ce{Cl+} formation, albeit with the atomic cation likely in a closed-shell singlet excited state.
    
    \item \textcolor{black}{\textbf{Double Ionization of \ce{CCl4}:}
    The recollision processes may have sufficient energy to not merely excite an electron in \ce{CCl4+}, but to ionize it further to \ce{CCl4^{2+}}, as has been observed for other species\cite{hao2022recollision,haan2007recollision}. The vertical double ionization energy of \ce{CCl4} leading to the ground state of \ce{CCl4^{2+}} is 28 eV, indicating that this is a potentially plausible outcome. Ground state  \ce{CCl4^{2+}} is a triplet, with both holes in the Cl 3p lone pair levels.  The equilibrium geometry of this state is a D$_{2d}$ symmetry form with a 1.75 {\AA} bond length (vs 1.77 {\AA} in neutral \ce{CCl4}). The computed C K-edge absorption features of this species are very similar to neutral \ce{CCl4} (as the holes are on nonbonding Cl levels and bond lengths are similar), indicating that it will be invisible to this probe.
    \newline It is thermodynamically favorable (by 4 eV) for this D$_{2d}$ form of \ce{CCl4^{2+}} to dissociate into \ce{CCl3+} and \ce{Cl+}. However, the barrier for this process is fairly large (0.70 eV) relative to the barrier for the C$_{2v}$ form of \ce{CCl4+} dissociating into \ce{CCl3+} and \ce{Cl} (0.17 eV). The lifetime of ground state \ce{CCl4^{2+}} should thus be much larger than the 90 fs lifetime observed for the SBCB forms of \ce{CCl4+}. 
    It therefore appears rather unlikely that any dication (if formed) contributes significantly to the observed C K-edge transient signal at $<$ 300 fs, as such species are likely to remain in the metastable D$_{2d}$ form, which is invisible to our probe. Excited dication states involving holes in the bonding or Cl 3s lone pair levels are unlikely, as no bright low energy spectral features (similar to C/D photoionization states of the cation) are detected. 
    }
\end{enumerate}




\textcolor{black}{To summarize, excited states of \ce{CCl4+} and even doubly ionized \ce{CCl4^{2+}} might reasonably be expected to form under experimental conditions. However, we find no evidence of species with holes in the $\sigma_{CCl}$ bonding levels or Cl 3s lone pairs, which would lead to prominent low energy features in the transient spectra. Species with holes in the Cl 3p lone pairs are plausible as excitations that fill such holes are dark with respect to our probe (especially at the C K-edge). However, the} observed features for the Cl formation channel appear to be adequately explained via just the X electronic state of \ce{CCl4+} alone, without need for any higher energy states. 
Higher vibrational states are however possible, though, and indeed quite likely. \textcolor{black}{The potential role of excited \ce{CCl4+} states (with Cl 3p lone pair holes) in forming \ce{Cl+} via a side channel is briefly discussed later.}

\section{Electric Field Effects Near Time Zero}
The presence of the pump electric field can cause additional signals in the $\Delta$OD data during the temporal overlap of pump and probe. The most common effect in strong-field experiments is a mixing of electronic states with dipoles along the electric field, referred to as the Stark effect\cite{krems2018molecules}. \textcolor{black}{A peak pulse intensity of $3\times10^{14}$ W/cm$^2$ corresponds to a root-mean-squared electric field strength of 0.05 atomic units (a.u.),} which is large enough to significantly affect the spectrum. In general, the applied field should break the triple degeneracy of the $8t_2^*$ level and lead to mixing with the $7a_1^*$ level. Calculations of neutral \ce{CCl4} with an applied field of 0.05 a.u. (at various orientations relative to the molecular axis) indicate that excitations to some of the initial $8t_2^*$ orbitals are blue-shifted in energy to $\sim 291.5$ eV, while transitions to the former $7a_1^*$ level are around 288.2-288.6 eV (which is now weakly absorbing due to mixing with the $8t_2^*$). Thus, the positive feature at 288.8 eV in the experimental spectrum likely arises from neutral \ce{CCl4} with antibonding levels perturbed by the electric field. 
This peak is observed in all pump intensity conditions, with larger shifts at higher intensities and smaller shifts moving away from time zero. The Stark effect also affects the spectrum of the ions, but the unionized species is present in greater amounts and hence is likely dominant in the observed spectrum.

\section{\label{section:ClPlusFormation} Additional Channel Leading to C\MakeLowercase{l}$^+$ Formation}

\FloatBarrier
\begin{figure}[b]
\includegraphics[max size={\linewidth}{\textheight/2}]{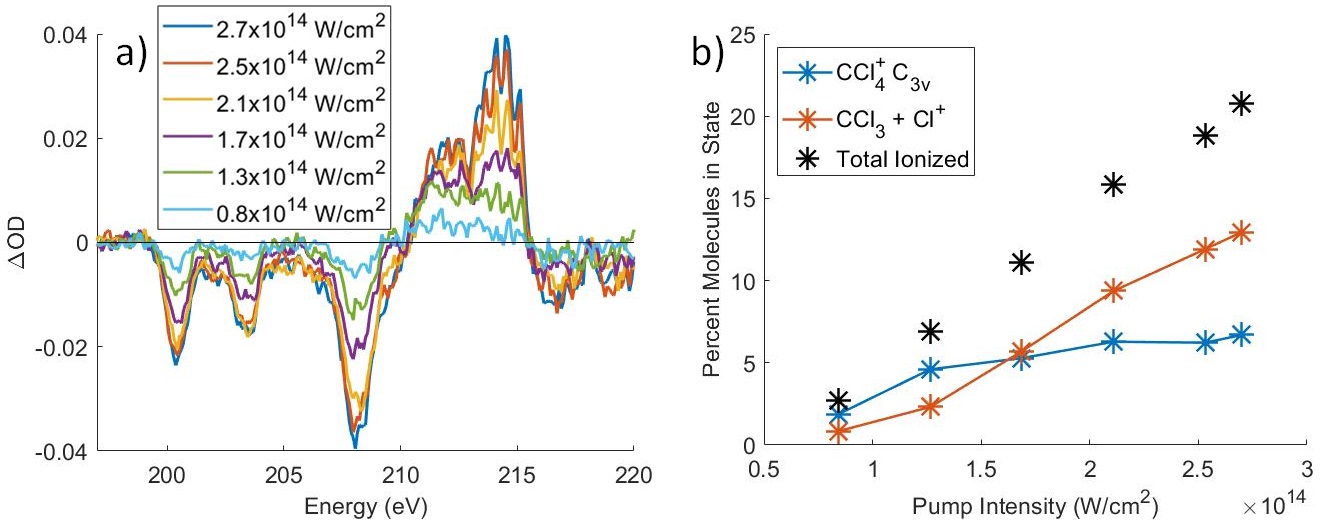}
\caption{\label{fig:figureFunctionOf} 
a: Spectral lineouts averaged from 190-230 fs as a function of increasing pump power, measured before the vacuum chambers and recombination mirror. Note that at high powers, the sharp lines at 214 eV corresponding to \ce{Cl+} become increasingly prevalent. b: Percentage of molecules in each channel between generation of atomic neutral Cl or cationic \ce{Cl+}. The X-axis uses estimated intensities of the pump pulse, but the relative intensities are accurate. At the powers and pulse duration used, the amount of the neutral channel plateaus quickly and has only slight increases with increasing power, whereas the \ce{Cl+} channel increases nearly linearly after $\sim1.3\times10^{14}$ W/cm$^2$. At the highest power, the relative amounts of Cl and \ce{Cl+} from \ce{CCl4+} dissociation are 35$\%$ and 65$\%$, respectively.}
\end{figure}

While the observed dissociation process to CCl$_3^+$ and Cl can be well explained by the intermediate forms and pathways discussed in the main manuscript, another channel that appears to result in neutral CCl$_3$ and cationic Cl$^+$ constitutes a significant part of the data. This channel is clearest at the chlorine L$_{2,3}$-edge where very sharp lines with similar widths to the atomic Cl lines are observed at 214 eV with less obvious components at 212.6 eV, shown in Fig. \ref{fig:figureFunctionOf}, which is the range that our ab initio calculations predict for the Rydberg states ($2p\to 3d,4s$) of Cl$^+$. These atomically sharp peaks begin to appear with a delay of 37$\pm$6 fs relative to the onset of the main cationic CCl$_4$ signal, and from that point, it shows a time to rise of 85$\pm$10 fs for the best signal-to-noise dataset. For the intensity scan, these dissociation times are found to be 70$\pm$40 fs, 94$\pm$94 fs, and 229$\pm$559 fs, for 2.8, 2.3, and 1.8 $\times10^{14}$ W/cm$^2$, respectively. This suggests the dissociation times of \ce{Cl+} have a slight dependence on pump intensity; although, uncertainties in the data do not allow a quantitative analysis. The delay of 37 fs is similar to the 23 fs delay of the 287.1 eV NBC signal to start appearing, which suggests that the time it takes for a Cl$^+$ atom to move far enough from the molecule to show atomic lines is about that time $\sim$37 fs, and that the nuclear motion may be similar to the case that forms neutral Cl. 

\begin{figure}
\includegraphics[max size={\linewidth}{\textheight/2}]{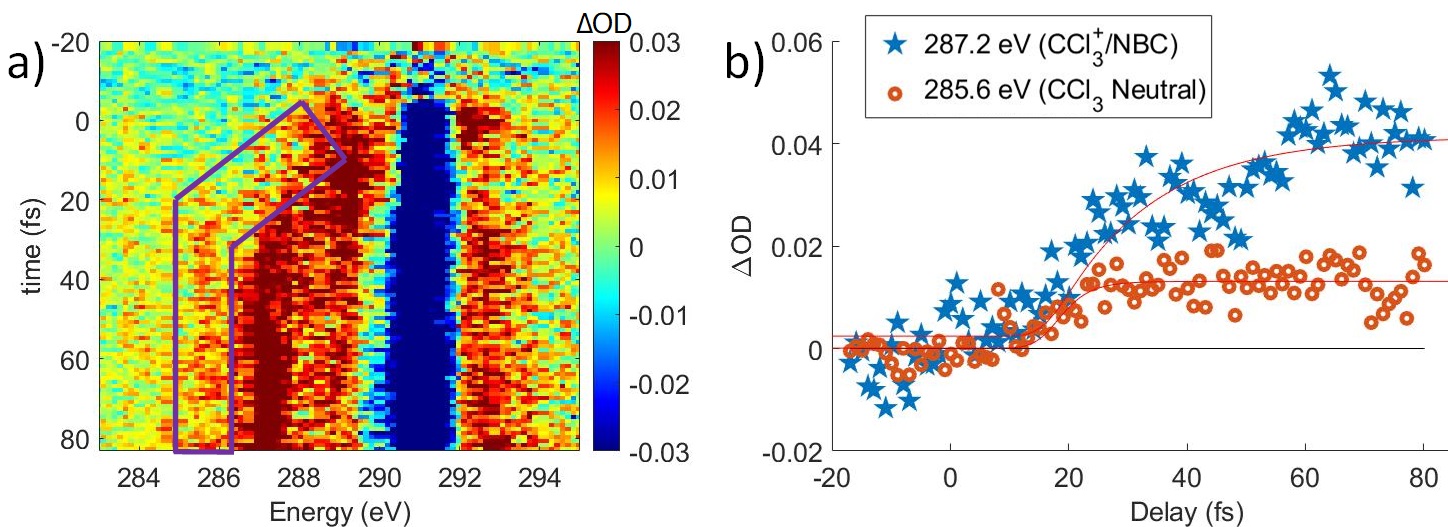}
\caption{\label{fig:ClPlusFormationCKedge} a: $\Delta$OD colormap at the C K-edge with colors saturated to better show a feature that splits from the main feature and continues to 285.5 eV.
b: Lineouts comparing the growth of the 287.2 eV \ce{CCl3+}/NBC feature to the 285.5 eV feature, which is assigned to neutral \ce{CCl3}. The NBC lineout has a delay of 23 $\pm$ 8 fs and a rise time of 50 $\pm$ 20 fs. The \ce{CCl3} lineout has a delay of 20 $\pm$ 10 fs and a rise time of 8 $\pm$ 6 fs.
}
\end{figure}

After this delay, there is the somewhat long exponential time of 85$\pm$10 fs for the lines to reach their maximum value, which may suggest the presence of a shallow energy minimum, possibly corresponding to a NBC complex, before complete dissociation. The \ce{Cl+} lines appear at much earlier times than \ce{Cl} lines for two possible reasons. Firstly, the 3d and 4s levels of the cation have greater valence character due to the positive charge, leading to stronger absorption that is harder to broaden into the baseline. Secondly, the associated \ce{CCl3} moiety for this pathway is neutral and thus plays less of a role in influencing the energy levels of \ce{Cl+} than \ce{CCl3+} does for \ce{Cl}.

\begin{figure}
\includegraphics[max size={\linewidth}{\textheight/2}]{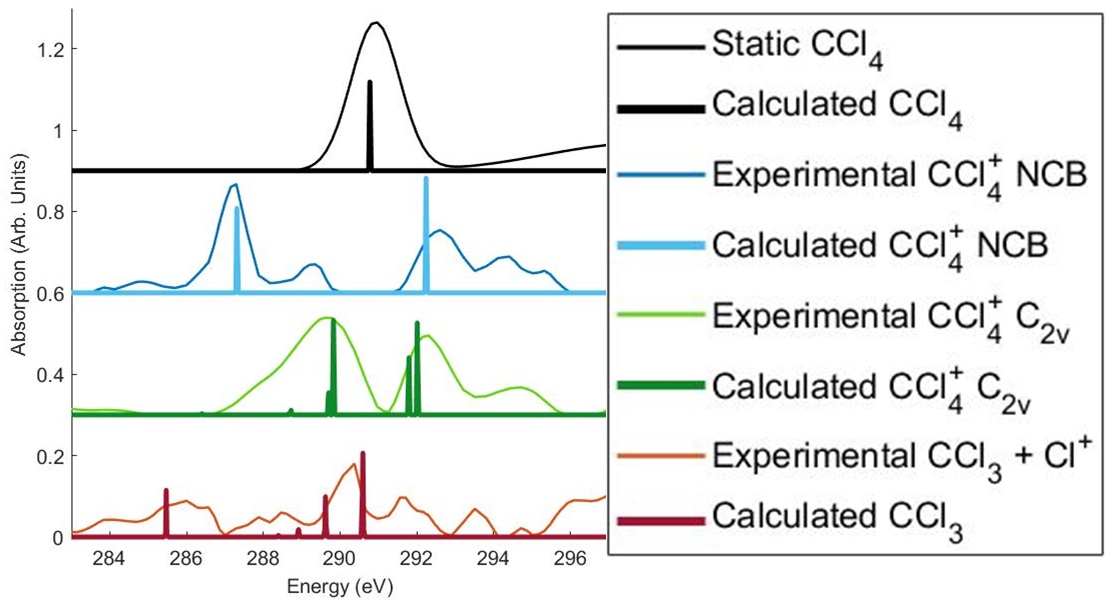}
\caption{\label{fig:spectrumOutputWithCCl3} A comparison between experiment and theory, including neutral \ce{CCl3}, SBCB form, and NBC shows good agreement with each.
}
\end{figure}

At the C K-edge, evidence of this channel can also be seen in a feature that appears at 285.6 eV, shown in figure \ref{fig:ClPlusFormationCKedge}. OO-DFT calculations show \ce{CCl3} is expected to have absorption at 285.5 eV. It shows a delay in appearance of 20 $\pm$ 10 fs, followed by a rise time of 8 $\pm$ 6 fs. The delay is very similar to the 23 $\pm$ 8 fs delay in the NBC formation, which was associated with the time required for the nuclear motion to break the covalent bond for the molecules that dissociated immediately. Again the similarity of the delay suggests similar nuclear motion along the pathways. However, the much shorter rise time suggests that only the molecules that dissociate immediately contribute to the \ce{Cl+} formation pathway.

The formation of Cl$^+$ is a much higher energy channel, with a final energy that is 4.8 eV higher than the Cl dissociation channel. Cl$^+$ formation has been observed in previous \ce{CCl4} ionization experiments from both electron impact\cite{Lindsay2004ElectronImpact} and single-photon ionization\cite{Kinugawa2002photoionization,Burton1993Photoionization400eV}. However, it was always found to be a minor channel, with an abundance of 0-5$\%$ around $\sim 20$ eV ionization energy. A larger percentage of \ce{Cl+}, $\sim$36$\%$, is observed in strong-field ionization\cite{geissler2007concerted}, and, in the power scaling shown in Fig. \ref{fig:figureFunctionOf}, we observe more than 65$\%$ forming, with the difference from other results\cite{geissler2007concerted} likely arising from shorter pulse durations and higher intensities. The high energy of this channel appears to suggest that it arises via an excited state, potentially one involving an electron excited from the Cl lone pair SALCs to the anti-bonding $\sigma^*_{\ce{CCl}}$ orbitals. Such a state would not be acessible under single-photon photoionization conditions. Occupation of a repulsive $\sigma^*_{\ce{CCl}}$ level in particular, can lead to rapid bond cleavage.

\section{Time Constant Changes with Varying Pump Intensity}
\FloatBarrier

\begin{figure*}
\includegraphics[max size={\textwidth}{\textheight}]{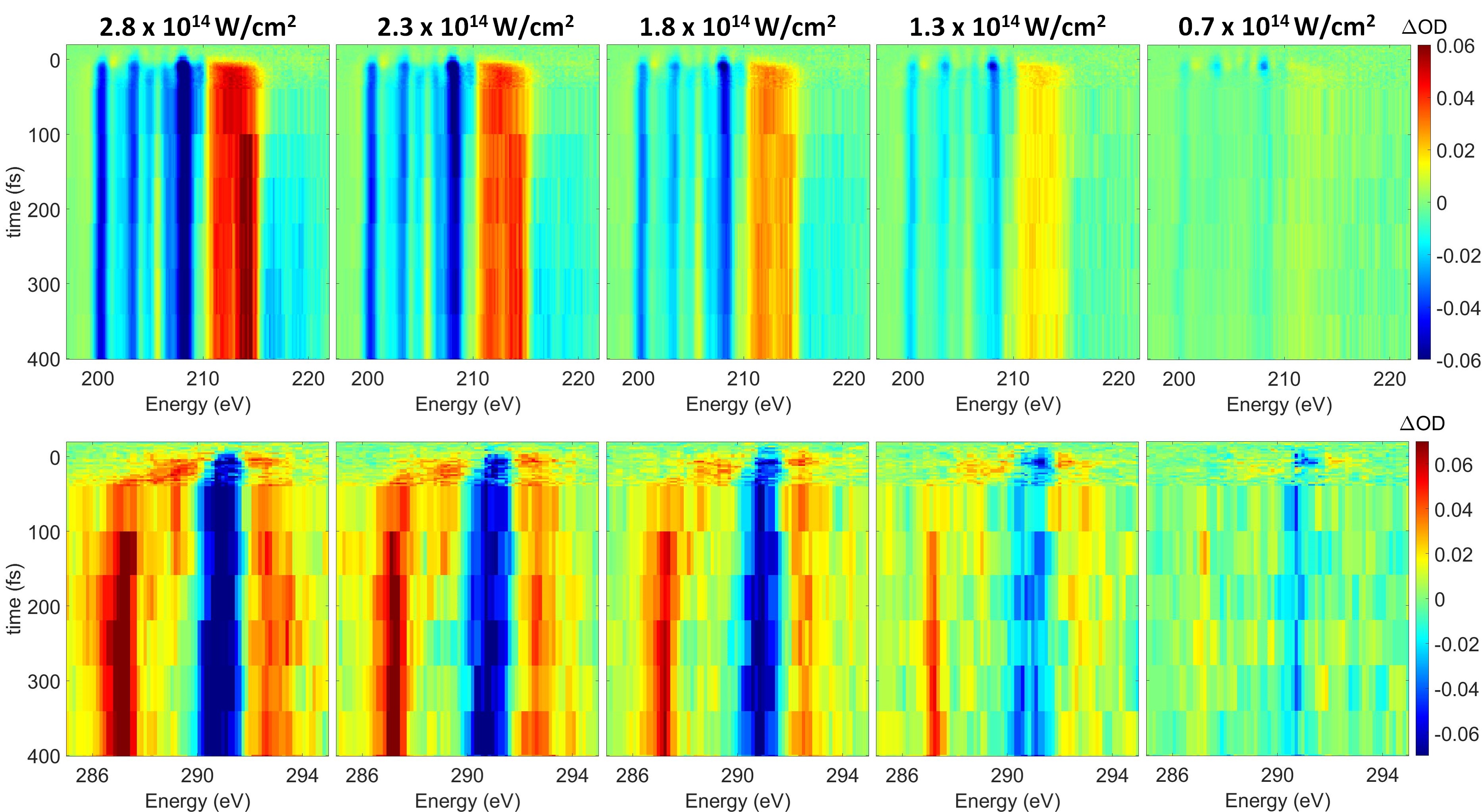}
\caption{\label{fig:dOD2022-01-30 Powerscan} Experimental data taken as a power scan. Top: Cl L-edge. Bottom: C K-edge. The estimated intensities of the pump pulse in each condition is shown.  Extracted lifetimes from fitting are given in Table \ref{table:Powerscan Time constants}.
}
\end{figure*}

In order to determine the dependence of the lifetimes measured on the pump intensity, a measurement was taken that varied the pump power. The parameters of the power scan are as follows: For the part near time zero, 2.5 fs steps are used, out to 40 fs. Then 60 fs steps are used to 400 fs. Powers used between these are directly comparable to each other, as the data was taken as simultaneously as possible. A scan from -20 to 400 fs ($\sim$6 min process was taken at one power, followed by a scan at the next power, and so on until a full scan was taken at each power. Then, another scan at the first power was taken, followed by the next, and so on for $\sim$8 hours total. The power dependence of signals were determined from this power scan. These data do not show a statistically significant difference in the T$_d$ $\to$ SBCB time or the SBCB $\to$ NBC time. The scan shows a possible difference in the lone, ionic Cl$^+$ formation time. The relative amounts of \ce{Cl+} and \ce{Cl} formation are also determined from this scan. The $\Delta$OD data are shown in Fig. \ref{fig:dOD2022-01-30 Powerscan}, and the fitted lifetimes are given in Table \ref{table:Powerscan Time constants}.

\begin{table}[]
\begin{tabular}{|l|l|l|l|}
\hline
Estimated Intensity       & T$_d$ $\to$ SBCB (fs) & SBCB $\to$ NBC (fs) & Atomic \ce{Cl+} rise at 214 eV (fs) \\ \hline
2.8$\times$10$^{14}$ W/cm$^2$ & 8 $\pm$ 4             & 90 $\pm$ 10         & 70 $\pm$ 40                         \\ \hline
2.3$\times$10$^{14}$ W/cm$^2$ & 5 $\pm$ 4             & 90 $\pm$ 10         & 90 $\pm$ 90                         \\ \hline
1.8$\times$10$^{14}$ W/cm$^2$ & 4 $\pm$ 5             & 90 $\pm$ 20         & 200 $\pm$ 600                       \\ \hline
1.3$\times$10$^{14}$ W/cm$^2$ & 6 $\pm$ 6             & 100 $\pm$ 20        & None observed                       \\ \hline
0.7$\times$10$^{14}$ W/cm$^2$ & 6 $\pm$ 6             & 90 $\pm$ 50         & None observed                       \\ \hline
\end{tabular}
\caption{\label{table:Powerscan Time constants}
Time constants for fits of the data in Fig. \ref{fig:dOD2022-01-30 Powerscan}. Powers are shown as estimated intensity. The atomic \ce{Cl+} rise time refers to the sharp spectral features at 214 eV and each rise follows a $\sim$37 fs delay, as described in section \ref{section:ClPlusFormation}. 
}
\end{table}

\FloatBarrier

\section{Calculated Geometries and Energies of Intermediates}

The exact structures optimized and used for calculations are given for each species as separate xyz files, with key structural parameters noted in Table \ref{table:IntermediateSpecificsSI}. Structures were optimized with the $\omega$B97M-V\cite{wB97MV} density functional and the aug-pcseg-3\cite{jensen2014unifying} basis set. Zero-point energies were found at the same level of theory. Relative ground state electronic energies at the optimized geometries were computed with CCSD(T)\cite{raghavachari1989fifth} extrapolated to the complete basis set (CBS) limit. The CBS extrapolation was done as follows:
\begin{align}
    E\left[\text{CCSD(T)/CBS}\right]=&E\left[\text{HF/CBS}\right]+E_{corr}\left[\text{CCSD(T)/CBS}\right]\\
    E\left[\text{HF/CBS}\right]\approx& E\left[\text{HF/aug-cc-pV5Z}\right]\\
    E_{corr}\left[\text{CCSD(T)/CBS}\right] \approx& \dfrac{64E_{corr}\left[\text{fc-CCSD(T)/aug-cc-pVQZ}\right]-27E_{corr}\left[\text{fc-CCSD(T)/aug-cc-pVTZ}\right]}{64-27}\notag\\
    &+E_{corr}\left[\text{CCSD(T)/aug-cc-pCVTZ}\right]-E_{corr}\left[\text{fc-CCSD(T)/aug-cc-pCVTZ}\right]
\end{align}
where $E_{corr}$ is the correlation energy, and fc stands for frozen-core (i.e. only correlation from valence electrons is computed). The frozen-core correlation energy is extrapolated to the complete basis set limit using the two point, cubic extrapolation formula from Ref \citenum{helgaker1997basis}. The Dunning aug-cc-pVnZ  (n=T,Q,5) basis sets\cite{dunning1989gaussian,kendall1992electron,woon1993gaussian} were used for this purpose.  Correlation arising from core-electrons is corrected for via the $E_{corr}\left[\text{CCSD(T)/aug-cc-pCVTZ}\right]-E_{corr}\left[\text{fc-CCSD(T)/aug-cc-pCVTZ}\right]$ term, which is the difference between all-electron and frozen-core CCSD(T) results for the aug-cc-pCVTZ basis set\cite{woon1995gaussian,peterson2002accurate}. 
\FloatBarrier
\begin{table}[htb!]
\begin{tabular}{|l|l|l|l|l|l|l|}\hline
             & Energy (eV) & r$_{C-Cl_1}$ ({\AA}) & r$_{C-Cl_2}$ ({\AA}) & 	$\measuredangle_{Cl_1,C,Cl_1}$ ($^{\circ}$) & $\measuredangle_{Cl_1,C,Cl_2}$ ($^{\circ}$) & $\measuredangle_{Cl_2,C,Cl_2}$ ($^{\circ}$) \\ \hline
\ce{CCl4+} T$_d$     & 0.57        & 1.76          &              & 109.5               &             &                \\ \hline
\ce{CCl4+} C$_{2v}$    & 0.18        & 1.70          & 1.82          & 115.64            & 86.11           & 112.90             \\ \hline
\ce{CCl4+} NBC C$_{3v}$    & -0.15        & 1.64         & 3.38          & 119.90             &  90.43           &              \\ \hline
\ce{CCl3+} and Cl & 0           & 1.64        &              &  120              &             &                \\ \hline
\ce{CCl4}         & -11.13      & 1.77        &              & 109.5               &           &               \\ \hline
\end{tabular}
\caption{\label{table:IntermediateSpecificsSI} The calculated CCSD(T)/CBS energies, bond distances, and angles are given for the intermediates identified in this experiment. The dissociation limit energy is set to 0. For species with two inequivalent types of chlorines, Cl$_1$ always refers to the one with the shortest bond distance from C. For fully equivalent species, the Cl$_2$ terms are omitted.}
\end{table}

\FloatBarrier

\section{Calculated Normal Modes}
The normal modes were found at the $\omega$B97M-V/aug-pcseg-3 level of theory. Lack of rotational symmetry due to the finite nature of the numerical grid used for evaluating DFT exchange-correlation integrals lead to slight differences between computed normal mode frequencies that should be degenerate in theory. For instance the asymmetric stretch of \ce{CCl4} is triply degenerate, but the computed frequencies range from $810.24-811.49$ cm$^{-1}$.

\begin{table}[htb!]
\begin{tabular}{lrrlrr}
\multicolumn{3}{c}{\ce{CCl4}}                                                          &  & \multicolumn{2}{c}{Covalently bonded \ce{CCl4+}}                        \\
Experiment\cite{shimanouchi1973tables}              & \multicolumn{1}{l}{Theory} & \multicolumn{1}{l}{Period} &  & \multicolumn{1}{l}{Theory} & \multicolumn{1}{r}{Period} \\
(in cm$^{-1}$)              & \multicolumn{1}{l}{(in cm$^{-1}$)  } & \multicolumn{1}{l}{(in fs)  } &  & \multicolumn{1}{l}{(in cm$^{-1}$)  } & \multicolumn{1}{r}{(in fs)} \\
\multicolumn{1}{r}{217} & 220.45                     & 151.31                     &  & 198.92                     & 167.69                     \\
                        & 220.50                     & 151.28                     &  & 236.59                     & 140.99                     \\
\multicolumn{1}{r}{314} & 319.29                     & 104.47                     &  & 292.70                     & 113.96                     \\
                        & 319.31                     & 104.46                     &  & 299.41                     & 111.41                     \\
                        & 319.42                     & 104.43                     &  & 329.51                     & 101.23                     \\
\multicolumn{1}{r}{459} & 476.19                     & 70.05                      &  & 482.55                     & 69.13                      \\
\multicolumn{1}{r}{776} & 810.24                     & 41.17                      &  & 534.72                     & 62.38                      \\
                        & 811.02                     & 41.13                      &  & 801.70                     & 41.61                      \\
                        & 811.49                     & 41.11                      &  & 926.68                     & 36.00                     
\end{tabular}
\caption{Normal mode frequencies of \ce{CCl4} and covalently bonded \ce{CCl4+} in cm$^{-1}$. The oscillation period (in fs) computed from theoretical frequencies is also reported. Experimental normal mode frequencies of neutral \ce{CCl4} are provided for comparison.}
\label{tab:my-table}
\end{table}
\FloatBarrier
\section{Calculated C K-edge Transitions and Assignments}

OO-DFT\cite{hait2021orbital}/SCAN\cite{SCAN} calculated energies of transitions and their assignments from the theoretical calculations are given for each species. Detailed protocols for running such calculations are described in Refs \citenum{hait2020highly} and \citenum{hait2020accurate}.  Excited state orbital optimization was done with the square gradient minimization (SGM\cite{hait2020excited}) and initial maximum overlap method (IMOM\cite{barca2018simple}) algorithms, for restricted open-shell and unrestricted calculations, respectively. The aug-pcX-2 basis\cite{ambroise2018probing} was used at the site of the core-excitation, and aug-pcseg-2\cite{jensen2014unifying} was used for all other atoms. Energies are not empirically shifted to align with experiment. 

\begin{table}[htb!]
\begin{tabular}{|lll|}
\hline
\multicolumn{3}{|l|}{Neutral \ce{CCl4}}                                                                                          \\ \hline
\multicolumn{1}{|l|}{Transition Energy (in eV)}   & \multicolumn{1}{l|}{Oscillator Strength} & Assignment        \\ \hline
\multicolumn{1}{|l|}{289.00} & \multicolumn{1}{l|}{6.5796E-08}      & $7a_1^*$ (C -Cl $\sigma^*$)      \\ \hline
\multicolumn{1}{|l|}{290.77} & \multicolumn{1}{l|}{0.07267}      & $8t_2^*$ (C -Cl $\sigma^*$)      \\ \hline
\multicolumn{1}{|l|}{290.77} & \multicolumn{1}{l|}{0.07267}      & $8t_2^*$ (C -Cl $\sigma^*$)      \\ \hline
\multicolumn{1}{|l|}{290.77} & \multicolumn{1}{l|}{0.07267}      & $8t_2^*$ (C -Cl $\sigma^*$)      \\ \hline
\end{tabular}
\caption{\label{table:calculated neutral} C K-edge energies for neutral \ce{CCl4} used in Fig \ref{fig:spectrumOutputWithCCl3}. The C $1s\to7a_1^*$ transition is forbidden, and does not appear in the experimental spectrum. The $8t_2^*$ level is triply degenerate, leading to three transitions (as shown).}
\end{table}

\begin{table}[htb!]
\begin{tabular}{|lll|}
\hline
\multicolumn{3}{|l|}{\ce{CCl4+} C$_{2v}$}                                                                                          \\ \hline
\multicolumn{1}{|l|}{Transition Energy (in eV)}   & \multicolumn{1}{l|}{Oscillator Strength} & Assignment        \\ \hline
\multicolumn{1}{|l|}{286.4}    & \multicolumn{1}{l|}{0.001031}      & SOMO (Cl 2p)                                 \\ \hline
\multicolumn{1}{|l|}{288.6474} & \multicolumn{1}{l|}{0.000738}      & C -Cl symmetric $\sigma^*$ (long bonds)      \\ \hline
\multicolumn{1}{|l|}{289.7012} & \multicolumn{1}{l|}{0.018316}      & C -Cl symmetric $\sigma^*$ (long bonds)      \\ \hline
\multicolumn{1}{|l|}{288.7216} & \multicolumn{1}{l|}{0.003823}      & C -Cl antisymmetric $\sigma^*$ (long bonds)  \\ \hline
\multicolumn{1}{|l|}{289.8263} & \multicolumn{1}{l|}{0.077694}      & C -Cl antisymmetric $\sigma^*$ (long bonds)  \\ \hline
\multicolumn{1}{|l|}{290.573}  & \multicolumn{1}{l|}{0.000134}      & C -Cl symmetric $\sigma^*$ (short bonds)     \\ \hline
\multicolumn{1}{|l|}{291.7935} & \multicolumn{1}{l|}{0.046993}      & C -Cl symmetric $\sigma^*$ (short bonds)     \\ \hline
\multicolumn{1}{|l|}{290.8488} & \multicolumn{1}{l|}{1.21E-07}      & C -Cl antisymmetric $\sigma^*$ (short bonds) \\ \hline
\multicolumn{1}{|l|}{292.0107} & \multicolumn{1}{l|}{0.075429}      & C -Cl antisymmetric $\sigma^*$ (short bonds) \\ \hline
\end{tabular}
\caption{\label{table:calculated C2v} C K-edge energies for the C$_{2v}$ form of \ce{CCl4+} used in Fig \ref{fig:spectrumOutputWithCCl3}. The C$_{2v}$ form has 2 inequivalent bond types with different characteristic $\sigma^*$ energies. The transition from the carbon 1s to the chlorine 2p SOMO is computed to be at 286.4 eV, but with very low intensity, and this transition is not observed in the experimental data.}
\end{table}

\begin{table}[htb!]
\begin{tabular}{|lll|}
\hline
\multicolumn{3}{|l|}{\ce{CCl4+} NBC / \ce{CCl3+}}                                                                \\ \hline
\multicolumn{1}{|l|}{Transition Energy (eV)}   & \multicolumn{1}{l|}{Oscillator Strength} & Assignment              \\ \hline
\multicolumn{1}{|l|}{291.07}   & \multicolumn{1}{l|}{4.40E-05}      & SOMO (Cl 2p hence CT)   \\ \hline
\multicolumn{1}{|l|}{286.0362} & \multicolumn{1}{l|}{7.71E-09}      & C 2p                    \\ \hline
\multicolumn{1}{|l|}{287.3108} & \multicolumn{1}{l|}{0.07711}       & C 2p                    \\ \hline
\multicolumn{1}{|l|}{289.5408} & \multicolumn{1}{l|}{4.24E-08}      & C-Cl $\sigma^*$ ($a_1$) \\ \hline
\multicolumn{1}{|l|}{290.8483} & \multicolumn{1}{l|}{1.57E-05}      & C-Cl $\sigma^*$ ($a_1$) \\ \hline
\multicolumn{1}{|l|}{291.0506} & \multicolumn{1}{l|}{1.51E-05}      & C-Cl $\sigma^*$ (E)     \\ \hline
\multicolumn{1}{|l|}{292.2435} & \multicolumn{1}{l|}{0.064844}      & C-Cl $\sigma^*$ (E)     \\ \hline
\multicolumn{1}{|l|}{291.069}  & \multicolumn{1}{l|}{1.50E-05}      & C-Cl $\sigma^*$ (E)     \\ \hline
\multicolumn{1}{|l|}{292.2341} & \multicolumn{1}{l|}{0.06492}       & C-Cl $\sigma^*$ (E)     \\ \hline
\end{tabular}
\caption{\label{table:calculated C3v} The transition energies from the carbon 1s for the global minimum NBC form of \ce{CCl4+} used in Fig \ref{fig:spectrumOutputWithCCl3}.}
\end{table}

\begin{table}[htb!]
\begin{tabular}{|lll|}
\hline
\multicolumn{3}{|l|}{\ce{CCl3}}                                                            \\ \hline
\multicolumn{1}{|l|}{Transition Energy (eV)} & \multicolumn{1}{l|}{Oscillator Strength} & Assignment              \\ \hline
\multicolumn{1}{|l|}{285.47}                 & \multicolumn{1}{l|}{0.038337}      & SOMO (C2p like)         \\ \hline
\multicolumn{1}{|l|}{288.3928}               & \multicolumn{1}{l|}{0.001316}      & C-Cl $\sigma^*$ ($a_1$) \\ \hline
\multicolumn{1}{|l|}{288.9165}               & \multicolumn{1}{l|}{0.006173}      & C-Cl $\sigma^*$ ($a_1$) \\ \hline
\multicolumn{1}{|l|}{289.6242}               & \multicolumn{1}{l|}{0.023365}      & C-Cl $\sigma^*$ (E)     \\ \hline
\multicolumn{1}{|l|}{290.5987}               & \multicolumn{1}{l|}{0.049872}      & C-Cl $\sigma^*$ (E)     \\ \hline
\multicolumn{1}{|l|}{289.6126}               & \multicolumn{1}{l|}{0.023154}      & C-Cl $\sigma^*$ (E)     \\ \hline
\multicolumn{1}{|l|}{290.5866}               & \multicolumn{1}{l|}{0.048831}      & C-Cl $\sigma^*$ (E)     \\ \hline
\multicolumn{1}{|l|}{291.4134}               & \multicolumn{1}{l|}{0.000343}      & Rydberg $a_1$           \\ \hline
\multicolumn{1}{|l|}{292.7127}               & \multicolumn{1}{l|}{0.000317}      & Rydberg $a_1$           \\ \hline
\end{tabular}
\caption{\label{table:calculated CCl3} The transition energies from the carbon 1s for neutral \ce{CCl3} used in Fig \ref{fig:spectrumOutputWithCCl3}.}
\end{table}

\FloatBarrier

\section{Comment on Quasiclassical Trajectory Calculations}
We had utilized quasiclassical trajectory (QCT) calculations\cite{karplus1965exchange} to model the short-time symmetry breaking of T$_d$ \ce{CCl4+}. These calculations assume nuclei to be classical particles and propagate them on the ground state electronic potential energy surface produced from $\omega$B97M-V/aug-pcseg-1. The calculations therefore are on a single electronic energy surface (i.e. are adiabatic). The ionization process is assumed to be Franck-Condon, with the cation initially being in the neutral \ce{CCl4} equilibrium geometry at $t=0$. All 9 normal modes of the molecule are given kinetic energy corresponding to the zero-point energy (ZPE) of neutral \ce{CCl4}. Each normal mode has two possible velocity directions, leading to $2^9=512$ possible trajectories. The pump pulse likely supplies even greater amounts of energy into the vibrational modes, but this is not explored. Indeed, all the trajectory calculations were run in the absence of the field, to provide only a zeroth-order picture of the short time dynamics of the system. Absence of the field effects in fact might slow down the rate of processes in QCT vs experiment. 

The QCT trajectories however are not particularly reliable at long times due to the well-known problem of ZPE leakage\cite{miller1989simple}. The nuclear motion is classical, and there is no guarantee that each mode continues to retain at least ZPE of energy. Indeed, a `leakage' is generally observed from high frequency  to low frequency modes over time, affecting the dynamics. Such ZPE leakage affects our QCT simulations in two ways. Firstly, formation of the noncovalently bonded complex from covalently bonded \ce{CCl4+} involves an asymmetric stretch in the latter, which has relatively high frequency compared to the bending modes. Energy leakage out of the stretching modes to lower frequency modes therefore would hinder noncovalent complex formation, leading to the covalently bonded form persisting longer in the QCT calculations. Secondly, the noncovalent complex forms have three very low frequency modes corresponding to motion of atomic Cl relative to the \ce{CCl3+} moiety, and spurious energy flow into these modes through ZPE leakage would lead to rapid dissociation of the complex into constituent fragments. The ZPE leakage problem in the QCT calculations therefore appears to disadvantage the noncovalent complex, both by slowing the formation rate as well as reducing the lifetime against dissociation. The long time behavior of QCT trajectories therefore are not likely to be reflective of the actual dynamics of the system.

\color{black}
\section{Frequency of molecule-molecule and molecule-ion collisions}\label{sec:collision}
The \ce{CCl4} was purchased as a liquid and was made gaseous by exposure to the vacuum, using a needle valve to regulate the pressure. The gas pressure of the \ce{CCl4} in the foreline was set to 12 mbar.  No heating elements were used, so the temperature should be less than or equal to room temperature in the sample cell. Therefore, the number density of \ce{CCl4} molecules is:
\begin{align}
    \dfrac{N}{V} &= \dfrac{p}{k_BT} = 2.9\times 10^{23} m^{-3}\\
\end{align}
Collision theory yields the frequency (per unit volume) of molecule-molecule collisions to be:
\begin{align}
    Z = (\pi R^2) v_{rel} \left(\dfrac{N}{V}\right)^2
\end{align}
where $\pi R^2$ is the cross-sectional area of the collision and $v_{rel}$ is the average relative velocity between particles.
The `lifetime' of a single molecule against collision with another molecule is then given by:
\begin{align}
    \tau_{MM}=\dfrac{1}{(\pi R^2) v_{rel} \left(\dfrac{N}{V}\right)}
\end{align}
From kinetic theory:
\begin{align}
    v_{rel}&=\sqrt{\dfrac{8\pi k_BT}{\pi \mu}}
\end{align}
where $\mu$ is the relative mass for the colliding particles. For \ce{CCl4} $\mu = \dfrac{151.8754}{2}$ a.m.u. or $6.3\times 10^{-26}$ kg. Consequently, $v_{rel}=408$ m/s. The cross-section radius $R$ is estimated to be twice the C-Cl bond-length (1.76 {\AA}) plus twice the van dar Waals radius of Cl (2 {\AA}). With this, we obtain a cross-sectional area of $1.8\times 10^{-18}$ m$^2$. All of these quantities collectively yield
$\tau_{MM}=4.7$ ns.
This indicates that collisions between molecular species in our experiment occurs in the timescale of nanoseconds, and is thus not relevant on the timescales observed here. All the studied processes can therefore be safely assumed to be unimolecular. 

A more challenging case is the possibility of collisions between molecular species and the electrons produced by strong-field ionization. Such electrons can be much faster, permitting more frequent collisions that might be of comparable timescales as the observed dynamics. We therefore attempt to find a lower bound for such collision lifetimes to determine if they are relevant. Since only about a quarter of the molecules are ionized at most (from Fig \ref{fig:figureFunctionOf} ), the electron density is at most $\dfrac{N}{4V}$. If we assume the electrons have an energy of 20 eV, then they will have a speed of $v_{elec}=2.6\times 10^6$ m/s. This is much larger than the thermal velocity of molecular species, and thus we can assume $v_{rel}\approx v_{elec}$ for such collisions. Literature estimates for electron-\ce{CCl4} collision cross-section is at most $A= 10^{-18}$ m$^2$ for such electron energies\cite{azevedo2000scattering}. With this, we obtain the mean lifetime of a molecular species for electron collision to be $\tau_{Me}=\dfrac{1}{A v_{elec} \left(\dfrac{N}{4V}\right)}=5.2$ ps. This is close to the upper limits of the greatest timescales measured in the experiment, but is large enough to indicate that molecule-electron collisions are unlikely to be relevant for any process other than the slow dissociation of the NBC to atomic Cl. Even that process is considerably faster (800 fs) at high pump powers, such that collision with ionized electrons should be irrelevant. Collisions with electrons may become relevant at lower pump powers where the NBC lifetime is longer, but it is worth noting that the ionized electron density will be even lower at those powers, leading to even less frequent collisions. We note that the very fast electrons that are most likely to cause collisions are also the ones that would be least affected by the electrostatic field of any cationic molecular species, indicating that the provided estimate for collision time is adequate even though it is based on neutral \ce{CCl4}. 
\color{black}
\section{Separation of Spectral Components from Experimental Data by Multivariate Fitting}

\begin{figure}[b]
\includegraphics[max size={\linewidth}{\textheight/2}]{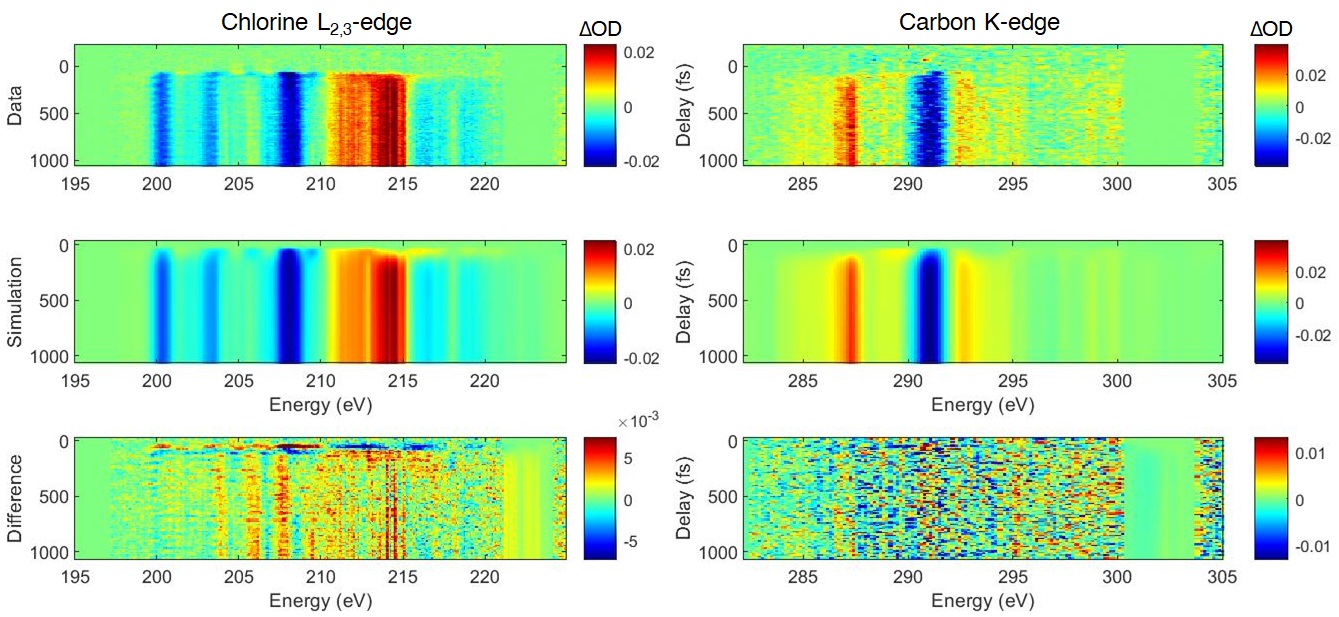}
\caption{\label{fig:exampleMultivariateFitting} 
An example of one dataset fitted by the multivariate fitting algorithm. Both the chlorine and carbon edges are shown in the two columns. The top row shows the data. The middle row shows the result of the fitting. The bottom row shows the difference between the data and the fitting (the top and middle rows). Note that the colorscale on the bottom row is zoomed by a factor of 2 to highlight errors. This shows only one dataset. For this fitting, 9 datasets are fitted simultaneously. The others are not shown for space considerations.
}
\end{figure}

After the data for each dataset is prepared and saved in files containing the $\Delta$OD for the chlorine and carbon edges, the static absorption generated by the pump-off spectra and a no gas spectrum from the start of the run, delay axis, and pixel-to-energy calibration axis. A dataset is the set of data taken continuously with no changes to the experiment, usually divided by being from different days, but also possible single days with multiple conditions. Slight differences in energy calibration and sample pressure are determined by comparing the static absorption of each dataset to a master static spectrum. Changes are made accordingly to make each dataset comparable to each other. These are all loaded into a single program to fit all of them to a model. An example of a finished fit for one dataset is shown in Fig. \ref{fig:exampleMultivariateFitting}.

The model follows the general assumption that 
\[{\Delta}OD = (P_1{\times}OD_1 +  P_2{\times}OD_2 + P_3{\times}OD_3 + ... + P_{static}{\times}OD_{static}) - OD_{static}\]
where, \(P_1 + P_2 + P_3 + ... + P_{static} = 1\) are coefficients denoting the percent of molecules in a particular state at a particular time and \(P_{static}\) is the percent that remain in neutral \ce{CCl4}. The optical density, OD, of each of the states is the goal of this fitting procedure. The number of states involved is able to be changed, based on the expected number of states. In this case, at least 3 are necessary to get a decent representation of the data, the C$_{2v}$, C$_{3v}$, and \ce{CCl3} + \ce{Cl+} forms.

The new absorptions, OD$_1$, OD$_2$, etc., are represented as parameters simply by absorptions spaced linearly between 2 energies. For example, the Cl L-edge used 201 points, and the C K-edge 81 points. These linear spacings were interpolated onto the nonlinear energy axis of each dataset. This representation of absorptions is very inefficient in terms of numbers of parameters, but it is exceptionally faster in converging to the global minimum. A set of pairs of absorption and the energy spacing from the previous point was also tried that needed about 1/3 of the number of parameters, but it took $\sim$40 times as many iterations to converge. The more chemically accurate representation with Gaussians corresponding to particular transitions was also attempted, and while it could represent the spectra with the fewest parameters, ~30 for the Cl L-edge and ~15 for the C K-edge, the parameter space was so riddled with local minima that it often would not change the parameters from their initial values. 

\begin{figure}[b]
\includegraphics[max size={\linewidth}{\textheight/2}]{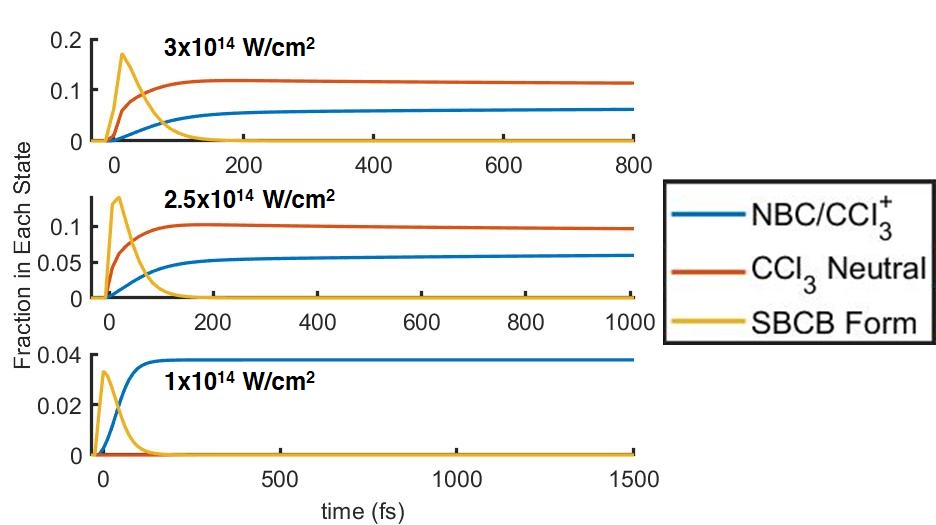}
\caption{\label{fig:exampleMultivariateTimeCurves} 
An example of the population curves used in a fitting are shown for 3 datasets in a 9 dataset fitting. The fraction in each state is defined such that 100$\%$ = 1, 20$\%$ = 0.2, etc. The 3 datasets represented have approximate intensity of the pump pulse of $3\times{}10^{14}$ W/cm$^2$, $2.5\times{}10^{14}$ W/cm$^2$, and $1\times{}10^{14}$ W/cm$^2$. The colors of the curves represent the same states in each plots, blue: NBC, red: \ce{CCl3} neutral after \ce{Cl+} dissociates, and yellow: SBCB form.
}
\end{figure}

The states are evolved in time to allow one state to become another and capture the dynamics of the system. When one state becomes another, its population is simply transferred into the absorption of the new one with no intermediate transition, because doing so would overly complicate the algorithm with minimal gain. The method of population fitting is to represent each state's \(P\) population as a function of time is with two single exponentials for rise and decay:
\[P_1(time) = (erf((x-c)/0.1 fs)+1) \times (-1*a*e^{(x-c)./(-1*\tau_{in})}+a) \times ((1-d)*e^{(x-c)/(-1*\tau_{out})}+d)\]
where \(c\) is a time offset to compensate for poorly decided time zeros, $\tau_{in}$ is the time constant for rise, $\tau_{out}$ is that for decay, \(a\) determines the maximum population, and \(d\) determines the final population of that state as a percentage of the maximum. An example of population curves resulting from a fit are shown in Fig. \ref{fig:exampleMultivariateTimeCurves}. The error function, erf, at the beginning is to set the population to zero before ionization and allow it to evolve after that. This function is chosen over subtraction of exponentials in order to be able to allow a population to finish at an arbitrary value, rather than 0, so that the same function can be used for initial, intermediate, and final states. Population curves are created for each state,
and the curves are then normalized to ensure \(P_1 + P_2 + P_3 + ... + P_{static} = 1\) at each time point. This is done to reduce the amount of local minima in the parameter space so that the fit can better find the global minimum, and this is the reason for several design decisions in this program. A potentially more accurate model might have been to assign k values between each state and calculate the populations numerically as a function of time; however, this would have required more time, which is anathema to use with a fitting procedure. 

The amount of total ionization for each dataset was determined by fitting to a single new state and minimizing the second derivative by transition energy of that state's absorption. This is similar to what is done in static spectral add back techniques\cite{Worner2017CF4,Ephshtein2020BenzenCation}, which will result in a percent ionization chosen that minimizes the contribution of the original static spectrum in the new spectra while ensuring that there is no negative absorption. When multiple new states are used, the ionization percent will tend towards 100$\%$ for the most intense cases, so it is instead held constant at the result determined by a single state. Minimization of the second derivative of the new states is also used when multiple new states are fitted to reduce overfitting between multiple states; however, it makes things like the Rydberg lines of Cl and \ce{Cl+} get smoothed over. 

In our data, we also see vibrations of $\sim$455 cm$^{-1}$, which corresponds to the symmetric stretch in neutral \ce{CCl4}, which will be further described in a later paper. These vibrations are taken into account in the program by moving the energies of the transitions to the molecular orbitals ($\sigma^*$ $7a_1^*$ and $\sigma^*$ $8t_2^*$). The static absorption is divided into Gaussians representing the transitions; and as a function of time, the centers of these Gaussians are moved in accordance with a sine function with the amount of movement dependent on the core-excited slope specific to that transition. 

The $\Delta$OD as a function of time is then calculated based on this model and is broadened by convolution with a Gaussian to account for the temporal resolution of our experiment. 

The model is subtracted from the data and fed into a minimization of least squares fitting algorithm, lsqnonlin in Matlab. The parameters that are common to all datasets are:
\begin{enumerate}
  \item OD of each new state
  \item Frequency of the neutral vibration
  \item Core-Excited state slopes for the vibration of each transition
  \item The lifetime constants, $\tau_{in}$, $\tau_{out}$, and \(d\), for each of the new states
\end{enumerate}
The parameters that are unique to each individual dataset are:
\begin{enumerate}
  \item Timing delay offset, \(c\)
  \item Population parameters, \(a\), for each state in each dataset
  \item Vibrational amplitude of the neutral
  \item Temporal broadening amounts
\end{enumerate}

\section{Method of Intensity Estimation by A\MakeLowercase{r} Ionization}
\begin{figure}[h]
\includegraphics[max height=\textheight/3,max width=\textwidth,keepaspectratio]{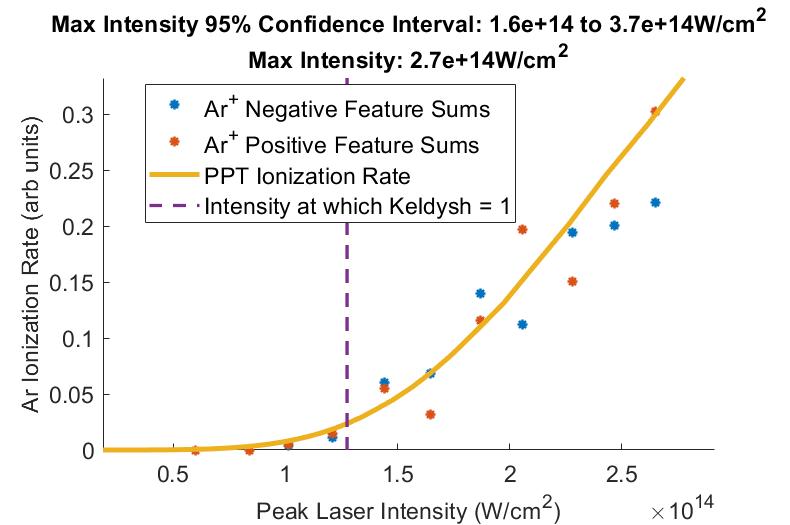}
\caption{\label{fig:Ar Ionization Comparison}
Internal calibration of the pump electric field intensity is done by comparison to the ionization rate of argon, using the PPT rate\cite{larochelle1998coulomb}. The intensity for which the Keldysh parameter is equal to 1 is also indicated, which is the intensity above which ionization is generally considered to be tunnel ionization, compared to multiphoton ionization\cite{wang2019identification}.}
\end{figure}

\FloatBarrier
A multiple measurements were taken at different powers on an Ar sample and the XUV features associated with \ce{Ar+} were added together into negative and positive features for t$\>$100 fs. The relationship between $\Delta$OD and ionization rate should be proportional, so the signals were linearly scaled until good agreement was found with the Perelomov, Popov and Terent’ev’s model (PPT) rate of the ionization of Ar\cite{larochelle1998coulomb} via a fit. Because of the linear scaling, the shape of the ionization rate and the power/intensity at which the rate changes are the largest factors in finding the correct intensity. The fit gives a rough approximation, but it gives an order of magnitude estimate for the electric field intensity of the pump of $2.7 \pm 1.1 \times{}10^{14}$ W/cm$^2$ at maximum power, shown in figure \ref{fig:Ar Ionization Comparison}. This is of the same order of magnitude as the peak intensity, $5\times{}10^{14}$ W/cm$^2$ given by calculation using 150 $\mu$J, 6 fs FWHM duration, and 65 $\mu$m FWHM spot size. The calculation assumes that the pulse is perfectly Gaussian in time and space, which is not necessarily true, especially since the beam is made collinear with an annular mirror. The calculation is expected to be an overestimate of the intensity, which is what is observed with comparison to Ar ionization.

\FloatBarrier

\section{Further Analysis of Distortion from Trajectory Calculations}
\FloatBarrier

\begin{figure}[h]
\includegraphics[max height=\textheight/2,max width=\textwidth,keepaspectratio]{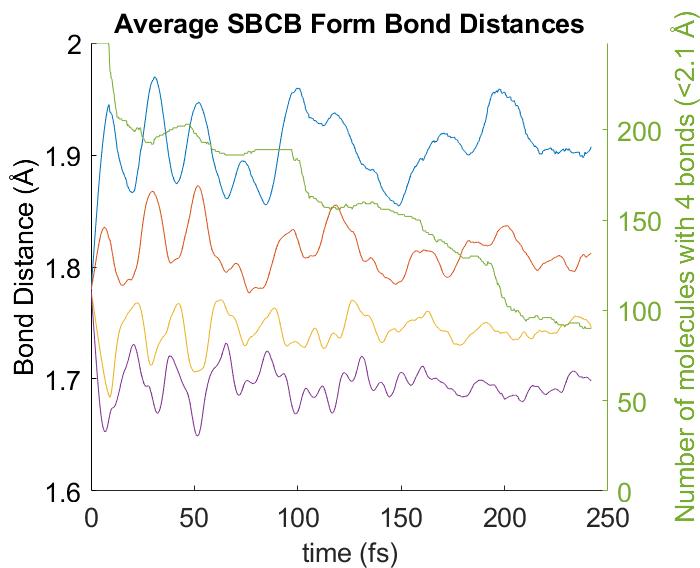}
\caption{\label{fig:Average SBCB Bond}
Average bond lengths for the \ce{CCl4+} ions that remain in the SBCB form are shown. The longest two bonds show opposite behavior to the two shortest bonds, consistent with a C$_{2v}$-like structure.}
\end{figure}

To better visualize the distortions in the SBCB form, the average of each bond length is shown in Fig. \ref{fig:Average SBCB Bond}. The lengths are binned together by their relative length at any given time. This ignores the particular identities of which chlorine the bond corresponds to and ensures that there will be no curve crossing. The averaging only includes ions that are still in the SBCB form, those that have 4 covalent C-Cl bonds, the number of which is shown in the green line on the right y-axis. It shows that the two long bonds extend outward from the ground state after ionization as the two short bonds contract. Following this, vibration-like behavior is observed as the two long and two short bonds mirror each other. 

In order to determine when the ion is expected to distort to the C$_{2v}$ minimum of \ce{CCl4+}, a variant of the root-mean-square deviation (RMSD) of atomic positions is used to compare the calculated positions at a given time from the trajectory calculations to the calculated C$_{2v}$ minimum structure. The exact formula used is:
\[RMSD = \sqrt{\frac{1}{N_{C-Cl}}\sum_{i=1}^{N_{C-Cl}}\delta_i^2 + \frac{1}{N_{Cl-Cl}}\sum_{i=1}^{N_{Cl-Cl}}\delta_j^2} \]
where $\delta$ is the difference between the inter-atomic distances for the trajectory distance and the C$_{2v}$ distance. This internal coordinate is used instead of the external XYZ coordinate to eliminate rotations and translations. A lower number indicates a structure more similar C$_{2v}$ structure. The initial value for the tetrahedral \ce{CCl4} is 0.44, and the first minimum appears at 26.6 fs with a value of 0.09. The RMSD becomes large when one of the C-Cl bonds break.

\begin{figure}[h]
\includegraphics[max height=\textheight/2,max width=\textwidth,keepaspectratio]{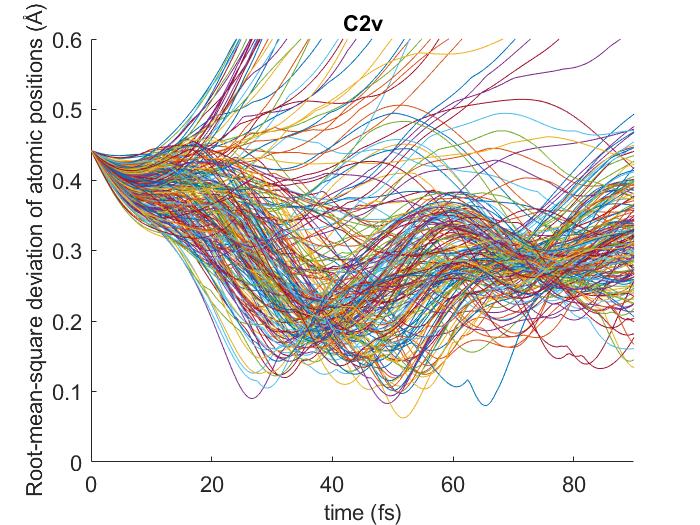}
\caption{\label{fig:C2v comparison}
Root-mean-square deviation of atomic positions from the local minimum C$_{2v}$ for all trajectory calculations. The smaller the number, the closer the structure is to C$_{2v}$. Trajectories show the ion can become the C$_{2v}$-like as early as 26.6 fs.}
\end{figure}

\FloatBarrier
\section{Additional CC\MakeLowercase{l}$_4^+$ Data}
\FloatBarrier

In the main text, several lineouts and timescales are taken using multiple datasets. These additional datasets comprise too many colormaps to show effectively in the main text, so they are shown here for completeness. \textcolor{black}{Also shown are lineouts taken at 2 different times from before and after the dissociation of the free Cl in Fig. \ref{fig:lineoutTimeComparisons}. These lineouts show that the changes in the spectra over the free Cl dissociation are minor, except in the area of 204-210 eV, where the atomic spectral features of the Cl become visible. Static and transient data at the Cl L$_1$-edge are shown in Fig. \ref{fig:L1EdgeStaticAndTransient}. The data for this figure come from the same measurement as the data for Fig. 2 in the main text. This edge was not analyzed due to the lower signal-to-noise ratio, as compared to the other edges.}

\begin{figure*}
\includegraphics[max size={\textwidth}{\textheight}]{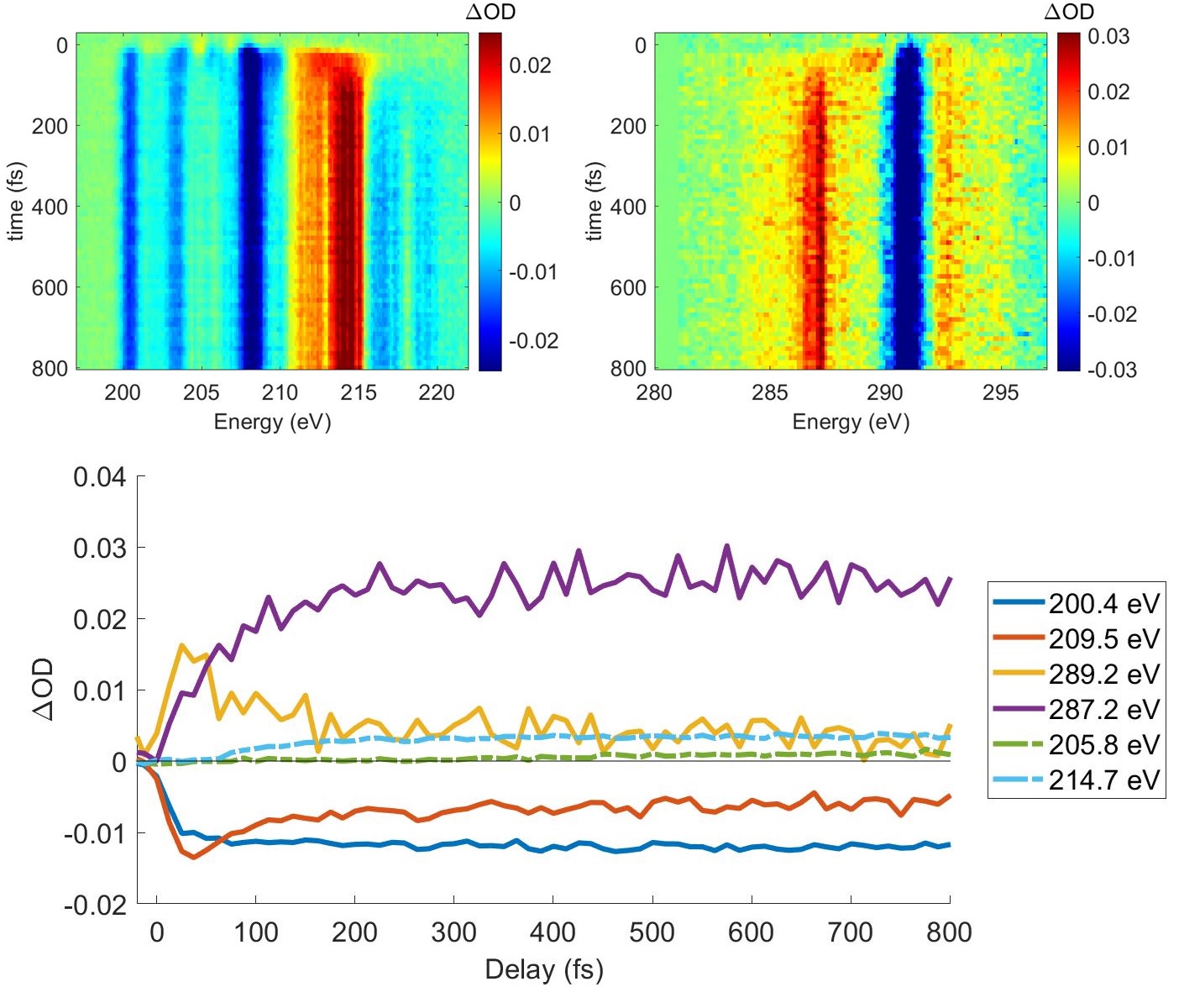}
\caption{\label{fig:dOD2021-06-18} 
Experimental data taken out to 800 fs with 12 fs steps. Maximum power is used here, $\sim$3$\times{}$10$^{14}$ W/cm$^2$. The dashed lines of 205.8 eV and 214.7 eV are taken after applying a high pass spectral filter to the $\Delta$OD data to isolate the parts that come from only the sharp spectral lines of the atomic Cl and \ce{Cl+}, respectively. These lineouts must be curated manually, as the energy calibration does not always correctly select the peaks, so the average energy (in the legend) may change depending on which of the multiple peaks were chosen. This is the dataset from which the $\sim$800 fs atomic Cl rise time was determined.}
\end{figure*}



\begin{figure*}
\includegraphics[max size={\textwidth}{\textheight}]{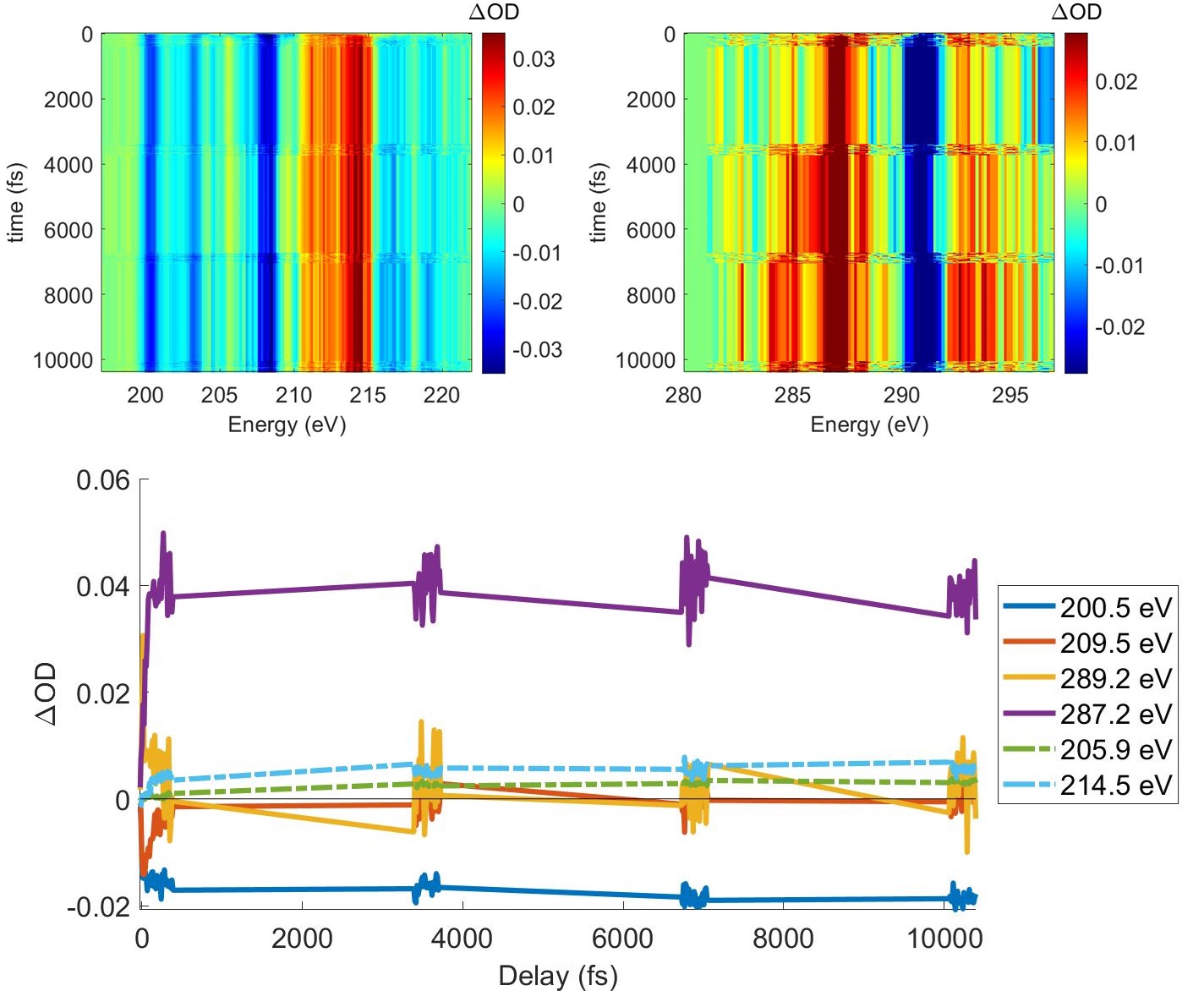}
\caption{\label{fig:dOD2021-07-08} Experimental data taken out to 10 ps. Data were taken in $\sim$300 fs blocks with 12 fs time steps with 4 blocks at 0 fs, 3.5 ps, 7 ps, and 10.5 ps in order to look at the vibrational differences between the regions. Near the maximum power is used here, $3\times{}10^{14}$ W/cm$^2$. These are the longest delays taken in the experiment. The comparison of the atomic Cl lines are taken from the average of these data at 220 fs and 3.5 ps. The lineouts also show that the signals do not change much after the time constants given in the text, up to 10 ps.}
\end{figure*}

\begin{figure*}
\includegraphics[max size={\textwidth}{\textheight}]{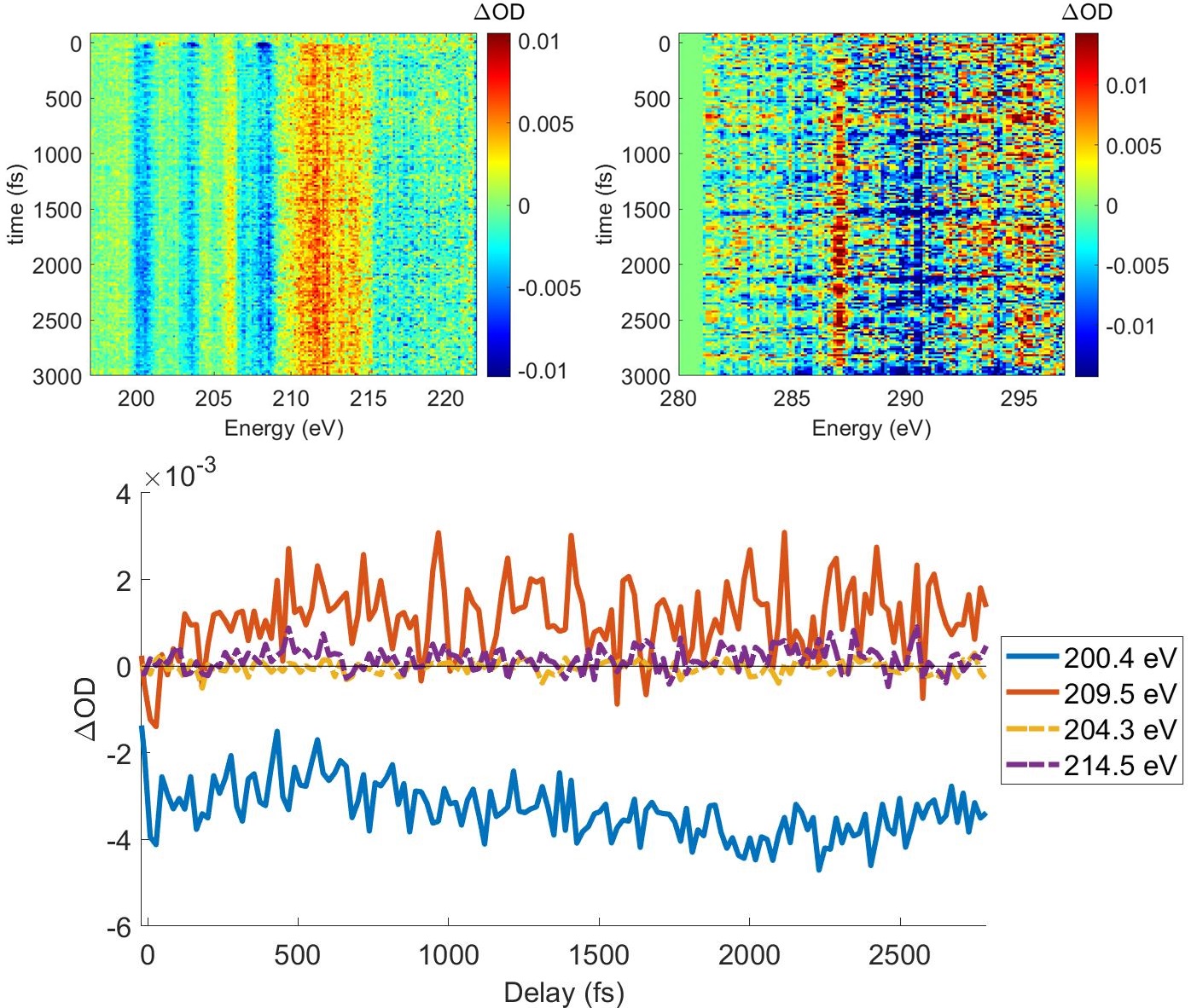}
\caption{\label{fig:dOD2021-08-16} Experimental data taken out to 3 ps. The focus in this scan was much looser than other datasets. Thus it represents a much lower power. No Cl formation in any significant amount is observed in this scan; although, the signals from \ce{CCl3+} are visible at 287.1 eV}
\end{figure*}

\begin{figure*}
\includegraphics[max size={\textwidth}{\textheight}]{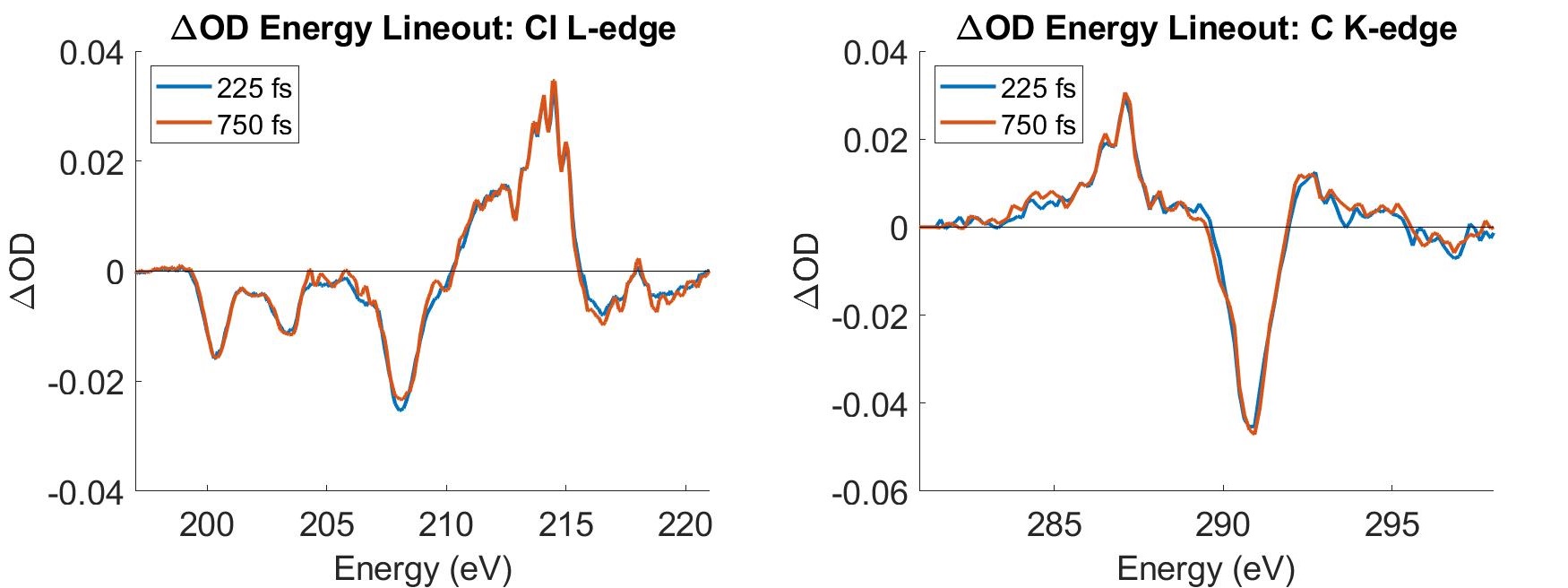}
\caption{\label{fig:lineoutTimeComparisons} \textcolor{black}{Experimental data from Fig. \ref{fig:dOD2021-06-18} with lineouts taken at 200-250 fs and 700-800 fs at the Cl L- and C K-edges. They show that there are very little changes to the $\Delta$OD spectra associated with the dissociation of the free Cl, other than the sharp spectral features from 204-210 eV.}
}
\end{figure*}

\begin{figure*}
\includegraphics[max size={\textwidth}{\textheight}]{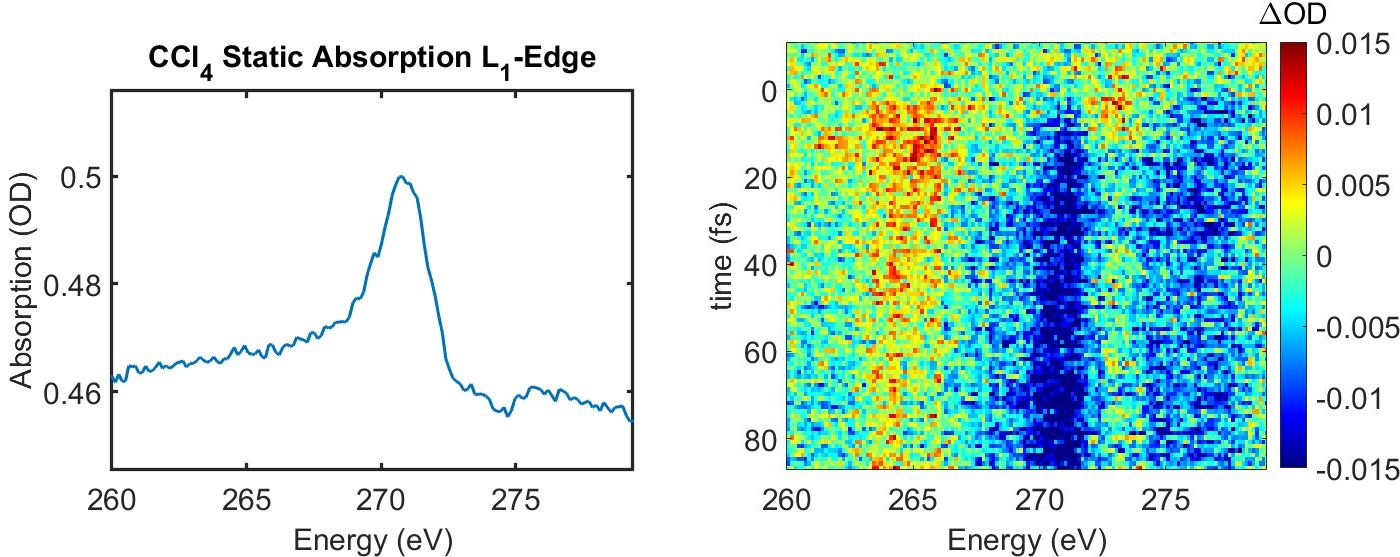}
\caption{\label{fig:L1EdgeStaticAndTransient} \textcolor{black}{Left: Static absorption of \ce{CCl4} at the Cl L$_1$-edge. Only a single, broad peak can be resolved. Right: Representative transient data at the Cl L$_1$-edge. The data for this figure come from the same measurement as the data for Fig. 2 in the main text. This was not analyzed due to the very low signal-to-noise ratio, as compared to the other edges.
}
}
\end{figure*}

\twocolumngrid
\section{References}
\bibliography{main}
\end{document}